%% file: 00-main.tex
\documentclass[manuscript, authorversion, nonacm]{acmart}

\AtBeginDocument{%
  }

\usepackage{xspace}
\usepackage{color, colortbl}
\usepackage{enumitem}
\usepackage[singlelinecheck=false]{caption}
\usepackage{tabularx}  % To have the table fill out the page
\usepackage{multirow}
\usepackage{multicol}
\usepackage{threeparttable}
\usepackage{float}
\usepackage[bottom]{footmisc}
\usepackage{subcaption}

\newcolumntype{P}[1]{>{\centering\arraybackslash}m{#1}}
\newcolumntype{R}[1]{>{\raggedleft\let\newline\\\arraybackslash\hspace{0pt}}m{#1}}
\newcolumntype{L}[1]{>{\raggedright\let\newline\\\arraybackslash\hspace{0pt}}m{#1}}
\newcolumntype{C}{ >{\centering\arraybackslash} m{4cm} }

\usepackage[capitalize,noabbrev]{cleveref}

\setlist[itemize]{align=parleft,left=0.5em..1.5em}

\newlist{req}{enumerate}{2}
\setlist[req,1]{label=RQ \arabic*:,ref= \textbf{\arabic*}, leftmargin=*}

\newlist{hyp}{enumerate}{2}
\setlist[hyp,1]{before=\itshape,font=\itshape, label=Hypothesis \arabic*:,ref= \arabic*, leftmargin=*}
\setlist[hyp,2]{before=\itshape,font=\itshape, label=Hypothesis \arabic*:,ref= \arabic*, leftmargin=*}

\begin{document}

% \title[Impact of Imperfect XAI]{The Impact of Imperfect XAI on Appropriate Reliance}
\title[Misinformation Effect of Explanations]{Don't be Fooled: The Misinformation Effect of Explanations in Human-AI Collaboration}

% ANOTHER SUGGESTION: Influential Factors of Explainable AI in AI-Assisted Decision-Making

\author{Philipp Spitzer}
\email{philipp.spitzer@kit.edu}
\orcid{0000-0002-9378-0872}
\affiliation{%
  \institution{Karlsruhe Institute of Technology}
  \country{Germany}
}

\author{Joshua Holstein}
\email{joshua.holstein@kit.edu}
\affiliation{%
  \institution{Karlsruhe Institute of Technology}
  \streetaddress{Kaiserstr. 89}
  \city{Karlsruhe}
  \country{Germany}
  \postcode{76133}
}

\author{Katelyn Morrison}
\email{kcmorris@cs.cmu.edu}
\orcid{0000-0002-2644-4422}
\affiliation{%
  \institution{Carnegie Mellon University}
  \city{Pittsburgh}
  \state{Pennsylvania}
  \country{USA}
}

\author{Kenneth Holstein}
\email{kjholste@andrew.cmu.edu}
\orcid{0000-0002-8369-3847}
\affiliation{%
  \institution{Carnegie Mellon University}
  \city{Pittsburgh}
  \state{Pennsylvania}
  \country{USA}
}

\author{Gerhard Satzger}
\email{gerhard.satzger@kit.edu}
\affiliation{%
  \institution{Karlsruhe Institute of Technology}
  \country{Germany}
}

\author{Niklas Kühl}
\email{kuehl@uni-bayreuth.de}
\orcid{0000-0001-6750-0876}
\affiliation{%
  \institution{University of Bayreuth}
  \country{Germany}
}

% \newcommand{\adam}[1]{\textcolor{magenta}{[#1 - adam]}}
% \newcommand{\katelyn}[1]{\textcolor{teal}{[#1 - katelyn]}}
% \newcommand{\philipp}[1]{\textcolor{purple}{[#1 - philipp]}}
% \newcommand{\violet}[1]{\textcolor{orange}{[#1 - violet]}}
% \newcommand{\niklas}[1]{\textcolor{green}{[#1 - niklas]}}

% \newcommand{\needscite}{\{\color{red}CITE\color{black}\} }

%%
%% By default, the full list of authors will be used on the page
%% headers. Often, this list is too long and will overlap
\renewcommand{\shortauthors}{Spitzer, et al.}

\begin{abstract}
Across various applications, humans increasingly use black-box artificial intelligence (AI) systems without insight into these systems' reasoning. To counter this opacity, explainable AI (XAI) methods promise enhanced transparency and interpretability. While recent studies have explored how XAI affects human-AI collaboration, few have examined the potential pitfalls caused by incorrect explanations. The implications for humans can be far-reaching but have not been explored extensively. To investigate this, we ran a study (n=160) on AI-assisted decision-making in which humans were supported by XAI. Our findings reveal a \textit{misinformation effect} when incorrect explanations accompany correct AI advice with implications post-collaboration. This effect causes humans to infer flawed reasoning strategies, hindering task execution and demonstrating impaired procedural knowledge. Additionally, incorrect explanations compromise human-AI team-performance during collaboration. With our work, we contribute to HCI by providing empirical evidence for the negative consequences of incorrect explanations on humans post-collaboration and outlining guidelines for designers of AI.

\end{abstract}

\begin{CCSXML}
<ccs2012>
<concept>
<concept_id>10003120.10003121.10011748</concept_id>
<concept_desc>Human-centered computing~Empirical studies in HCI</concept_desc>
<concept_significance>500</concept_significance>
</concept>
<concept>
<concept_id>10010147.10010178</concept_id>
<concept_desc>Computing methodologies~Artificial intelligence</concept_desc>
<concept_significance>500</concept_significance>
</concept>
   <concept>
       <concept_id>10010147.10010178.10010224.10010225</concept_id>
       <concept_desc>Computing methodologies~Computer vision tasks</concept_desc>
       <concept_significance>500</concept_significance>
       </concept>
   <concept>
\end{CCSXML}

\ccsdesc[500]{Human-centered computing~Empirical studies in HCI}
\ccsdesc[500]{Computing methodologies~Artificial intelligence}
\ccsdesc[500]{Computing methodologies~Computer vision tasks}

\keywords{Human-AI Collaboration, Explainable AI, XAI for Computer Vision}

%   \includegraphics[width=\textwidth]{Figures/teaserfigure1.pdf}
%   \caption{Ornithologists and citizen scientists collaborate with the same imperfect artificial intelligence (AI) systems. However, citizen scientists have a different level of expertise than ornithologists, which could cause them to under-rely on the AI advice, such as in this case with the Canada Warbler. In this example, the example-based explanation, which was generated based on a content-based image retrieval technique, shows a different bird species than the Canada Warbler. This incorrect explanation can cause non-experts, such as citizen scientists, to under rely on AI even when the prediction is correct. Experts may be more confident with their own knowledge, thus correctly classifying the bird despite the incorrect explanation. 
% %   % \katelyn{Thoughts on this figure?? also I used DALLE to create the two people -- do we need to cite that somehow? What are the current ACM policies on this?}
%    }
% %   % \violet{I think we might want to replace the figure with a pair of more diverse birders?}
%   \Description{}

%%
%% This command processes the author and affiliation, and title
\maketitle

\input{01-introduction}
\input{02-relatedworks}
\input{03-theoretical-framework}

\input{04-studydesign}
\input{05-results}

\input{06-discussion-limitations}

\input{07-conclusion}

%% identification of the section in the article metadata and the
%\begin{acks}

% individually name the people who helped share our study
% ask Luke and the other person about naming them
% Add undergrad to this
% @Adam and Niklas, which grants should we be acknowledging for funding / supporting this work?
%%\end{acks}
\section*{Acknowledgements}

Generative AI tools were utilized throughout this work. Specifically, ChatGPT, Claude and Github Copilot were employed to generate code for visualizations. Additionally, ChatGPT, DeepL Write, and Grammarly were used to enhance the writing quality of tutorials and explanations provided to participants during the experiments, as well as to improve the language across all sections of this paper.
\bibliographystyle{ACM-Reference-Format}
\bibliography{references}

\newpage
\appendix
\input{08-appendix}

\end{document}

%% file: 01-introduction.tex
\section{Introduction}
\label{sec: introduction}

\textit{Imagine you are studying for an art history exam and must know how to distinguish two architectural styles. You seek advice from an online artificial intelligence (AI) assistant for explanations to differentiate the two styles. Despite being plausible and providing you with correct architectural styles, the AI's explanation is incorrect. Yet, you learn from these incorrect explanations, and as a consequence, your understanding and capability to distinguish the architectural styles is impaired! You fail your exam.}

%Imagine you're preparing for an art history exam where you need to distinguish between two architectural styles. You seek advice from an artificial intelligence (AI) assistant to classify them. While the AI correctly identifies the architectural styles in examples you show it, its explanations focus on architectural elements that are not actually characteristic of those styles. By learning from these misleading explanations, you develop an incorrect understanding of what makes each style unique! You fail your exam.
%

%What might sound like an accidental event in an arbitrary situation in fact presents a crucial research discourse in human-computer interaction (HCI)  with respect to human-AI collaboration for decision-making \cite{morrison2024impact}. 
%With AI increasingly affecting various decision-making domains--from everyday applications like product recommendations to critical sectors such as healthcare, legal services, and finance \cite{monteith2024artificial, goldstein2023generative}--also the interaction with this technology undergoes fundamental changes. 
Recent technological advancements have significantly bolstered both the capabilities and the adoption of AI \cite{bommasani2021opportunities, dwivedi2023opinion}. For instance, large language models (LLMs) can assist in high-stakes classification tasks, such as categorizing legal cases based on case descriptions, by providing initial advice along with an explanation, even when users have no prior expertise in legal analysis \cite{ma2024leveraging}. In the medical field, AI has been used to accurately classify retinal disorders, leading to early diagnosis and providing critical support to ophthalmologists \cite{mukhtorov2023endoscopic}.
%In the medical field, generative AI can support individuals with a first diagnosis by recommending treatments for complaints \citep{el2023chatgpt}.
However, as these AI technologies become more sophisticated, especially with the use of generative AI, the complexity and opacity of supported decision-making processes increase. This presents new challenges in ensuring that these systems are still transparent and interpretable for humans \cite{rudin2022interpretable}. This increasing complexity necessitates a deeper understanding of the underlying factors that impact human-AI interaction, especially in scenarios where human oversight is critical \cite{amershi2019guidelines, sterz2024quest, cabitza2023painting}. Regulatory frameworks, such as the EU AI Act~\cite{EU_AI_Act_2021}, mandate human oversight to ensure AI systems operate ethically, legally, and safely. This underscores the importance of human-computer interaction (HCI) research to develop methods and tools that facilitate effective collaboration between humans and AI \cite{shneiderman2020human}. Ensuring that AI is not only accurate but also explainable is vital to foster appropriate reliance and complementary team performance \cite{schemmer2023appropriate, hemmer2023human}.
The need for explainability in AI has been widely acknowledged, leading to substantial advancements in both research and practice \cite{silva2023explainable, arrieta2020explainable}. In the financial sector, for instance, eXplainable AI (XAI) is used in credit scoring and fraud detection, targeting improved auditing, regulatory compliance, and user trust \cite{cirqueira2021towards}.

Despite these advancements, the XAI literature has barely picked up on an important potential pitfall in human-AI collaboration scenarios: incorrect explanations \cite{morrison2024impact, lakkaraju2020fool, kayser2024fool} for accurate AI advice. These incorrect explanations may impair humans \textbf{post-collaboration} as they compromise their \textit{procedural knowledge}---their ability to perform specific tasks on their own--- and their \textit{reasoning}---their ability to derive conclusions based on their understanding. Exploring the effect of incorrect explanations on procedural knowledge and reasoning is crucial, as regulations \cite{EU_AI_Act_2021} or performance expectations \cite{hemmer2023human} might require humans to complement AI capabilities with their domain knowledge and, therefore, maintain the ability to perform tasks on a superior performance level. This is particularly critical in high-stakes domains like healthcare, finance, and legal settings, where flawed decisions can have severe consequences \cite{rudin2019stop}. The preservation of human procedural knowledge becomes even more essential in scenarios where AI advice is initially provided but later removed, such as in AI-based teaching settings \cite{spitzer2023ml, spitzer2024transferring}, as humans need to maintain their ability to perform tasks independently when AI support is unavailable.

Understanding the impact of incorrect explanations in AI-assisted decision-making, even when the AI advice is correct, is essential for designing more effective collaborations between humans and AI. By identifying how and why incorrect explanations impact humans not only \textit{during}, but in particular \textit{post-collaboration} with AI, we can develop effective strategies to mitigate these effects \cite{chen2022interpretable, lakkaraju2020fool, ehsan2019automated, ribeiro2016should}. 
Thus, we ask the following research questions (RQs):

\begin{req}[leftmargin=1.06cm, labelindent=0pt, labelwidth=0em, label=\textbf{RQ\arabic*}:, ref=\arabic*]
    %\item How can we measure the impact of incorrect explanations on humans' procedural knowledge and reasoning in AI-assisted decision-making? \label{rq1}
    \item How do incorrect explanations for correct AI advice impair humans' procedural knowledge in AI-assisted decision-making?\label{rq1}
    \item How do incorrect explanations for correct AI advice impair humans' reasoning in AI-assisted decision-making?\label{rq2}
    \item How do incorrect explanations for correct AI advice affect the human-AI team performance?\label{rq3}
\end{req}

To address these questions, we first synthesize how to measure the impact of incorrect explanations in AI-assisted decision-making based on previous research. We pre-registered our hypotheses on AsPredicted.org\footnote{We attached an anonymized copy of the pre-registration to the submission of this article.}. Through an online study with 160 participants, we examine how incorrect explanations influence humans' procedural knowledge (\textbf{RQ \ref{rq1}}) and reasoning (\textbf{RQ \ref{rq2}}) in a task to classify architectural styles of buildings. We chose this classification task because AI, and especially LLMs, are increasingly being used nowadays not only to generate content but also to support human reasoning and decision-making \citep{hadi2024large}. Thus, they provide advice to humans in scenarios where no prior knowledge is needed (e.g., patients seeking medical advice) to help them interpret and draw conclusions from complex data. Such settings remain crucial as they provide a controlled environment (clear ground truth and measurable outcomes) to study how humans are impacted by incorrect explanations.
In this task, we provide participants with different types of AI support: no support at all, AI advice without explanations, AI advice with correct explanations, and AI advice with incorrect explanations. We measure the effects on their task performance during and after collaboration to derive the impact on their procedural knowledge and qualitatively analyze their understanding of the task to derive their reasoning ability. Additionally, we analyze the effect on the human-AI team performance (\textbf{RQ \ref{rq3}}).

The findings of our study reveal a \textit{misinformation effect} in AI-assisted decision-making: incorrect explanations significantly impair humans, resulting in a notable decline in their procedural knowledge once they have to perform the task autonomously. We also find that humans' reasoning is impaired when they receive incorrect explanations. In fact, we also observe a negative effect during the  collaboration in our study: human-AI teams perform worse when the AI provides incorrect explanations, curtailing the complementary benefits of this collaboration. 
These findings underscore the potential dangers of incorrect explanations and highlight the importance of developing robust and reliable explanatory support for humans.
%Interestingly, we find the human cognitive load has an impact on this effect.

In summary, our study makes several contributions to the field: first, it fills a critical gap in the literature by examining the impact of incorrect explanations on humans in AI-assisted decision-making. Second, it identifies and measures the misinformation effect within AI explanations and outlines its negative repercussions on humans---a decline in procedural knowledge and reasoning. Lastly, our research extends existing knowledge on the interplay between AI and human decision-makers, providing insights into the hazards of explanations on the human-AI team.
Overall, this paper sheds light on the potential pitfalls of incorrect explanations and their implications for humans post-collaboration. With this exploratory work, we hope to contribute to the development of concepts and hypotheses helping to advance theoretical knowledge in XAI and AI-assisted decision-making as well as to advance further the design of more effective AI-based decision support systems.

%% file: 02-relatedworks.tex
\section{Background}
\label{sec:relatedwork}

With the rapid advancement of AI and its integration into diverse decision-making processes, XAI has emerged as a critical technique for enhancing transparency and assistance to help decision-makers understand AI’s reasoning \cite{alufaisan2021does, cabitza2024never}. Especially with applications based on generative AI (like Open AI's ChatGPT or Anthropic's Claude), explanations are being generated in natural language that provide human-understandable support for the AI's response \cite{singh2024rethinking, zytek2024llms}. Previous research on human-AI collaboration has predominantly focused on how AI explanations influence human decision-making \cite{subramanian2024designing, schemmer2022influence}. For example, \citet{hemmer2021human} examine the factors that impact human-AI team performance, suggesting that XAI can foster complementary collaboration between humans and AI. Albeit the positive effects, studies also highlight how XAI can impair the collaboration between humans and AI \cite{miller2019explanation, ehsan2021expanding, binns2018fairness}. However, the understanding of the negative consequences of XAI is still limited \cite{mohseni2021multidisciplinary}, especially empirical evidence for the negative impact of incorrect explanations is missing. In \Cref{tab: litreview}, we sort recent works in AI-assisted decision-making according to the correctness of AI advice and explanations. The table shows the under-explored topic of incorrect explanations in HCI. We review recent literature highlighting the risks and limitations of collaborative settings between humans and AI, motivating the need to explore further how incorrect explanations impair humans.

\begin{table}[htbp!]
\caption{HCI literature investigating impacts of the correctness of AI advice and explanations on human-AI collaboration.}
\Description{The table shows in a two times two matrix how previous work contributed to correct/incorrect AI advice and correct/incorrect explanations.}
\label{tab: litreview}
\centering
\begin{tabular}{m{2.7cm} P{2.4cm}P{2.4cm}} 
\hline
 & \textbf{ Correct AI explanations} & \textbf{Incorrect AI explanations} \\ 
\hline
\textbf{Correct AI advice} & \cite{schemmer2022influence,hemmer2023human, schoeffer2024explanations, lai2019human, adhikari2019leafage, hase2020evaluating, ribeiro2018anchors, van2021evaluating, zhang2020effect, yeung2020sequential}& \cite{morrison2024impact, cabitza2024explanations, kayser2024fool}\\ 
\hline
\textbf{Incorrect AI advice} & \cite{sadeghi2024explaining, vicente2023humans, kim2023help, cau2023effects, buccinca2021trust, alufaisan2021does, ehrlich2011taking, chen2023understanding, cabitza2023rams, kocielnik2019will} & \cite{papenmeier2019model, morrison2024impact, lakkaraju2020fool, kayser2024fool} \\ 
\hline
\end{tabular}

\end{table}

\subsection{Incorrect AI Advice in Human-AI Collaboration}

Research in HCI has explored the effects of incorrect AI advice on human-AI collaboration \cite{kocielnik2019will, bansal2019beyond, yin2019understanding, vicente2023humans}. \citet{kocielnik2019will} investigate how such incorrect AI advice influences user satisfaction and acceptance. In their study, they demonstrate how interventions like accuracy indicators or performance control can maintain users' perception and trust even when the AI-based scheduling assistant provides wrong advice.  

Next to scheduling assistants, several recent studies have investigated how programmers collaborate with Copilot, an AI programming assistant that is not always accurate (e.g., \cite{vasconcelos2023generation, barke2023grounded, dakhel2023github}). For instance, \citet{vasconcelos2023generation} focus on an AI supporting code completion. As the AI can make errors in producing the code, they convey the uncertainty of Copilot's outputs by highlighting code sections that are most likely to be edited by the programmer. Their findings reveal that this approach helps programmers to arrive at solutions more quickly. Additionally, \citet{dakhel2023github} conclude that, while GitHub Copilot is a valuable tool for expert programmers, non-expert programmers should exercise caution when using it due to the potential for errors.
Other studies further examine how different levels of expertise among programmers influence their interaction with AI-generated code \citep{barke2023grounded}. Their findings suggest that while expert programmers can effectively navigate and correct AI-generated errors, novice programmers often struggle, leading to decreased efficiency and increased frustration.
While previous research shows the impact of incorrect AI advice on humans with different levels of prior domain knowledge, \citet{vicente2023humans} run a study on medical diagnosis to investigate how humans are influenced during collaboration with an AI that provides erroneous advice. They show that humans inherit the same errors as the AI, thus impairing their ability to conduct the task themselves.
%Moreover, \citet{yang2019unremarkable} analyze the role of trust in human-AI collaboration, particularly in scenarios where AI occasionally provides incorrect advice. Their findings suggest that initial trust can facilitate smoother interactions but that repeated exposure to incorrect advice significantly erodes trust.
Building on this, previous research also shows how non-expert users are influenced by AI systems providing inaccurate advice \citep{schemmer2023towards}. They find that users with limited domain expertise were prone to overreliance on the AI's outputs, resulting in decision errors. Contrarily, \cite{vasconcelos2023explanations} show that explanations can reduce overreliance by introducing a cost-benefit framework. They outline that explanations to be effective need to diminish the costs of verifying AI advice.

This review of prior research illustrates the potential negative impact of incorrect AI advice on decision outcomes and highlights how people's existing knowledge shapes their ability to interpret and respond to AI output. Building on these findings, the next section explores the role of incorrect explanations and how they influence humans' decision-making when supported by AI.

\subsection{Incorrect Explanations in Human-AI Collaboration}

Next to the AI advice, the understanding of how incorrect explanations can impact AI-assisted decision-making is limited. Only a few studies investigate the effect of explanations' incorrectness \cite{morrison2024impact, cabitza2024explanations, lakkaraju2020fool, papenmeier2019model}.
A recent study in HCI shows that not only incorrect advice but also incorrect explanations have the potential to deceive decision-makers \cite{morrison2024impact}. \citet{morrison2024impact} explore the negative impacts of incorrect explanations on humans' decision-making behavior. They extend the conceptualization of \citet{schemmer2023appropriate} by the explanation dimension and explore, in a bird classification study, how the correctness of explanations impacts humans' reliance on AI. They show that incorrect explanations can deceive decision-makers who possess no prior domain knowledge, leading to inappropriate reliance behavior.
\citet{cabitza2024explanations} explore the effects of explanations in a logic puzzle task. They also show that if advice and the accompanying explanation do not align, humans are misled ultimately resulting in inappropriate reliance behavior. \citet{papenmeier2019model} conduct another study in AI-assisted decision-making with incorrect explanations. They study how explanations affect humans' trust in identifying offensive tweets. Similarly, \citet{lakkaraju2020fool} find that incorrect explanations can affect humans' trust in AI by investigating their effects in law and criminal justice use cases.

These studies collectively underscore the complex dynamics of human-AI interaction, especially how incorrect advice and incorrect explanations affect humans' reliance behavior on AI. However, the HCI field lacks a deeper understanding of the effects of such impaired collaboration scenarios on humans themselves. Especially for scenarios in which not the advice but the explanation for the decision-maker---intended to foster interpretability---is incorrect. Studies like \citet{morrison2024impact} and \citet{cabitza2024explanations} build a promising starting point to inform HCI researchers and practitioners of the downsides of incorrect explanations. However, we still do not know anything about the impact of these AI shortcomings on humans' ability to perform the tasks autonomously (procedural knowledge) and to conclude about the underlying domain (reasoning) \textbf{post-collaboration}. 

%% file: 03-theoretical-framework.tex
\section{Theoretical Development}
\label{sec:theoretical-development}

In the evolving field of HCI, understanding the impact of AI explanations on decision-making has become critical. Explanations can serve as a bridge between AI and humans, influencing trust, reliance, and collaboration \cite{schemmer2023appropriate, hemmer2023human, schoeffer2024explanations}. This work investigates how incorrect explanations, when paired with accurate AI advice, can mislead humans, potentially impairing their procedural knowledge and reasoning capabilities. In \Cref{sec:relatedwork}, we present several works that investigate the impact of incorrect AI advice and incorrect explanations in decision-making. Building on prior research \citep{morrison2024impact, cabitza2024explanations, papenmeier2019model}, we derive several hypotheses grounded in related psychology and human behavior research to study the impact on humans' knowledge.

Research in behavioral science distinguishes between declarative knowledge (the ``know-what'') and procedural knowledge (the ``know-how'') in decision-making \citep{herz1999role}. While both types of knowledge are interconnected, our study focuses on procedural knowledge as we want to determine the impact of incorrect explanations on humans' downstream task performance. Humans' ability to complete tasks effectively remains paramount, especially where AI support is not perfectly reliable or may not always be available \citep{laux2023institutionalised}. %Furthermore humans must maintain the capability to make decisions independently \citep{kumar2023developing}. 
By measuring task performance, we can capture procedural knowledge \citep{mccormick1997conceptual, nahdi2020conceptual} and the extent to which humans effectively apply their understanding in practice \citep{clark2016regulation}.
Exposure to incorrect explanations can significantly distort both declarative and procedural knowledge. As procedural knowledge is particularly vulnerable to cognitive disruptions \citep{sweller1988cognitive}, incorrect explanations might increase cognitive load and impair the acquisition of procedural knowledge. \citet{chi2014icap} provide crucial insights into this phenomenon, demonstrating how cognitive engagement can be systematically undermined by misleading information. 
Thus, we assume that incorrect explanations for correct AI advice impair humans' procedural knowledge. We hypothesize:

\begin{hyp}[resume, wide, leftmargin=0cm, labelindent=0pt, labelwidth=0em]
\item Incorrect explanations for correct AI advice lead to lower procedural knowledge.
\label{hyp1}
\end{hyp}

While our study primarily centers on procedural knowledge, we also assess participants' reasoning abilities to explore how individuals articulate the cognitive principles underlying their decision-making processes \citep{johnson1994sources}. This enables us to examine the extent to which incorrect explanations influence both task performance and conceptual understanding.
Prior research has demonstrated the profound impact of misleading information on cognitive processes, revealing significant impairments in memory and comprehension \citep{loftus1978semantic, soon2018fake}. Critically, these cognitive distortions persist even after subsequent correction attempts \citep{ecker2011correcting, kendeou2013updating} demonstrating the effect of misinformation on cognition and their potential to impair reasoning \cite{ecker2011correcting}. In the context of AI-assisted classification tasks, these prior findings suggest that incorrect explanations are likely to disrupt reasoning ability in classification tasks.

\begin{hyp}[resume, wide, leftmargin=0cm, labelindent=0pt, labelwidth=0em]
\item Human reasoning is impaired by incorrect explanations for correct AI advice.
\label{hyp2}
\end{hyp}

\citet{seeber2020machines} emphasize the importance of communication between humans and AI for effective collaboration. They argue that misunderstandings or unclear communication can lead to disruptions in team performance, particularly when AI systems provide incorrect or unclear explanations. Furthermore, \citet{dzindolet2003role} investigate how trust in automated systems is impacted by incorrect feedback, leading to reduced team performance. Similarly, \citet{eiband2018bringing} provide insights into how human-AI collaboration is affected by the transparency of AI systems, suggesting that incorrect advice can hinder effective teamwork. Lastly, \citet{bansal2021does} explore the dynamics of human-AI interaction, showing that incorrect AI advice alongside explanations can create friction in collaboration, leading to poorer outcomes. Incorrect explanations in AI-assisted decision-making can impair the collaboration, leading to inappropriate reliance on AI \cite{morrison2024impact}.  
With prior research showing the relationship between humans' reliance behavior on AI and the human-AI team performance \cite{schemmer2023appropriate}, we assume that the human-AI team performance drops when humans are provided with incorrect explanations. Thus, we extend prior research by directly assessing the impact of incorrect explanations for correct AI advice and hypothesize:

\begin{hyp}[resume, wide, leftmargin=0cm, labelindent=0pt, labelwidth=0em]
\item Incorrect explanations for correct AI advice lead  to a lower human-AI team performance.
\label{hyp3}
\end{hyp}

%% file: 04-studydesign.tex
\section{Methodology}
\label{methodology}
In this section, we describe our methodology to assess how incorrect explanations for correct AI advice influence humans \textit{during} and \textit{post} collaboration with an AI. We set up an online study and investigated how participants performed in a visual classification task on an architectural dataset. We outline the task domain, the study design, the recruitment of participants, the development of the AI, and finally, the metrics that we use to assess our RQs. Before we ran the study, we pre-registered the study on AsPredicted.org to report our hypotheses, our treatments, our planned analyses, and our exclusion strategy. An anonymized copy of the pre-registration is provided in the supplemental materials. 

\subsection{Task Domain}

To analyze the impact of explanations, we chose a task where most people have little experience, i.e., have no expert-level knowledge and typically cannot perform well themselves initially: the classification of the architectural style of buildings. This task represents various real-world scenarios in which AI is utilized: seeking advice and explanations for a task that is either unknown or humans might not possess enough knowledge to solve the task confidently. Such scenarios in real-world use cases range from providing medical diagnoses for patients to generating code for people with no background in programming. Thus, the study of how humans integrate incorrect explanations into their decision-making likely extends to more complex generative AI interactions.
We use the established dataset of \citet{xu2014architectural} containing images of buildings across 25 different architectural styles. In close discussion with architecture researchers of the local university, we chose three architectural styles that share similar features and are not easily distinguishable: Art Nouveau, Art Deco, and Georgian Architecture. For each architectural style, we selected 30 images that clearly represent the features of each architectural style and, thus, are appropriate instances for our study. We further made sure that the buildings were centered in each image and cropped all irrelevant information in the images, like other buildings.

\subsection{Study Design}
\label{study-design-section}

Our research questions target the understanding of the impacts of incorrect explanations in AI-assisted decision-making on human procedural knowledge and reasoning and the resulting human-AI team performance. To address them, we employed a study combining between- and within-subjects design: between subjects, we analyze the impact of different AI support. Within subjects, we observe this impact in different stages of decision-making: before, during, and after collaboration with the AI. The study was approved by the university's institutional review board. 

\begin{figure}[!htp]
  \centering
  \includegraphics[width=\linewidth]{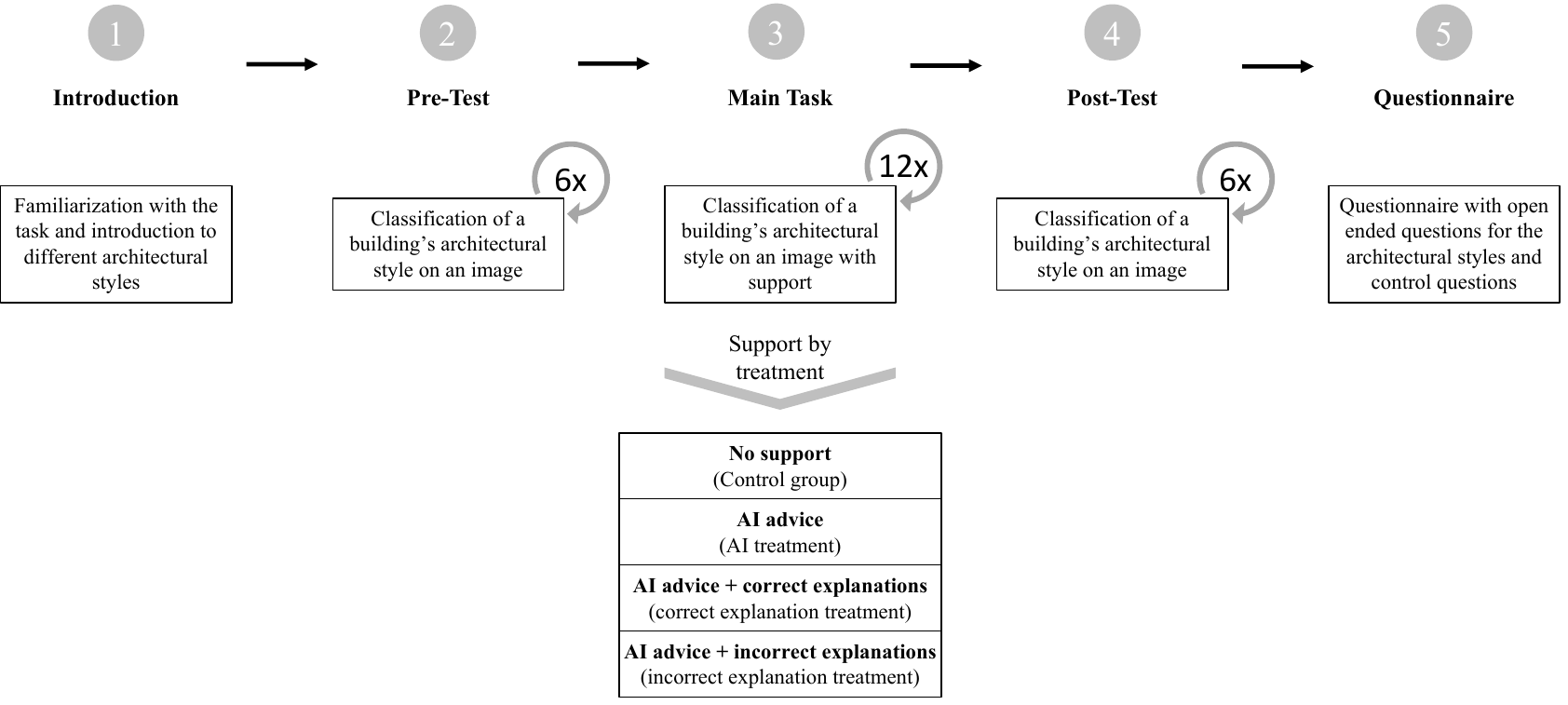}
  \caption{The study design is outlined in five different parts: in part (1), participants were introduced to the study. In part (2), participants had to classify six images as a pre-test. In part (3), participants were randomly assigned to a treatment and classified twelve images. In part (4), participants classified six images as post-test without support. Participants had to complete a questionnaire in the final part (5).}
  \Description{The figure shows the study procedure divided into five parts: The introduction, the pre-test, the main task, the post-test and the questionnaire.}
  \label{studydesign}
\end{figure}

The online study was divided into five different parts (see \Cref{studydesign}): in part (1), the participants had to give their consent to participate and were introduced to the study and its procedure. They also received context information about the three different architectural styles with a description of their main characteristics (see \Cref{tab: descriptions} in \Cref{sec: appendix}). Additionally, we included two attention checks, one of them in part (1): we asked participants what their task would be throughout the study, and they could select from three options. In part (2), a pre-test assessed participants' task performance in classifying architectural styles of buildings: they were each shown six images randomly drawn from a bucket of 18 pre-selected images of the dataset. We ensured that each class was balanced for each participant (two images per class) and that the order in which the classes were shown to participants was varied. By doing so, we minimized the risk of inducing biases in the order of images. It also ensured that our results were not dependent on the difficulty of pre-selected images. Moreover, the pre-test allowed us to check that participants in each treatment did not differ in their prior expertise regarding the task. 
Following the pre-test, participants were randomly assigned to a treatment in the main task in part (3), distinguishing the type of AI support received:

\begin{itemize}
    \item \textit{Control group:} Participants did not receive any AI support
    \item \textit{AI treatment:} Participants received only the AI's classification
    \item \textit{Correct explanation treatment:} Participants received the AI's classification and a correct explanation
    \item \textit{Incorrect explanation treatment:} Participants received the AI's classification and an incorrect explanation
\end{itemize}

\begin{figure}[!htp]
\centering
\begin{subfigure}{0.48\textwidth}

  \centering
  \includegraphics[width=\textwidth]{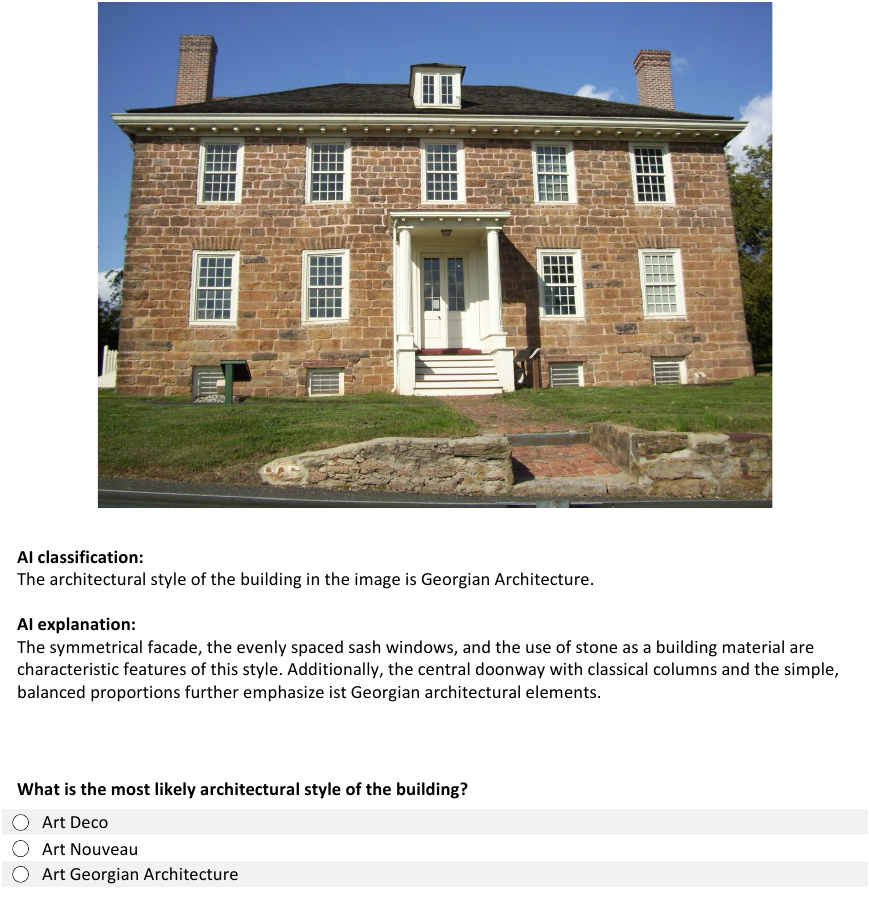}
    \captionsetup{justification=centering}

  \caption{Correct explanation treatment.}

\end{subfigure}
\begin{subfigure}{0.48\textwidth}
  \centering
  \includegraphics[width=\textwidth]{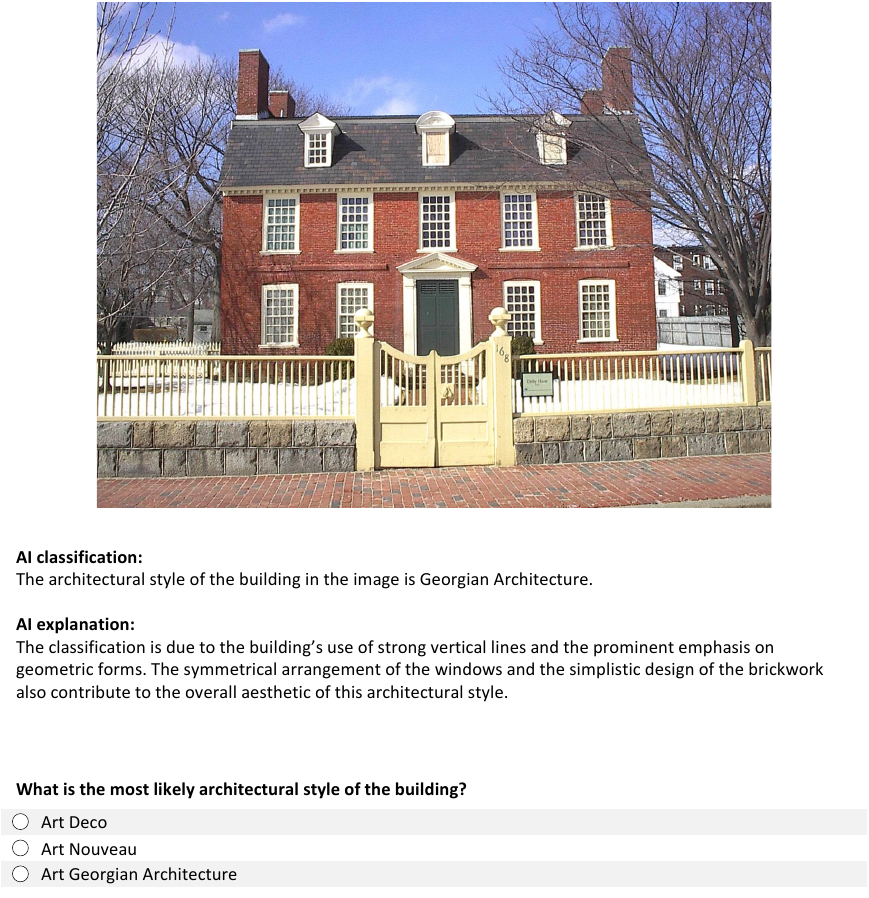}
    \captionsetup{justification=centering}

  \caption{Incorrect explanation treatment.}

  \centering
\end{subfigure}

    \centering
\captionsetup{justification=raggedright,singlelinecheck=false}
  \caption{Instances shown to participants in the main task with correct explanations (left) and incorrect explanations (right).  }
  \Description{The figure shows two examples of instances of the main task in the correct explanation treatment on the left and of the incorrect explanation treatment on the right.}
  \label{fig: example}
\end{figure}

By distinguishing the groups in this way, we can infer the effect of incorrect explanations on participants' procedural knowledge and reasoning. After the assignment, participants had to classify buildings' architectural styles on twelve images (see as an example \Cref{fig: example}). All twelve images were randomly drawn from a bucket of 30 pre-selected images balanced in classes. Similar to the pre-test, each participant was provided with four images of each class, and the order was randomized. Each image was displayed on a separate page in the study. During the main task, participants had the option to click on a button to show the context information for the architectures' characteristics and verify the support they received. 
In part (4), the post-test,  participants of each treatment had to classify six different images without support. Similar to the pre-test, the images were drawn from another bucket of 18 pre-selected images balanced in classes. Each participant was shown two images of each class on separate pages in randomized order.

To prevent participants from saving time by just randomly clicking through the study, participants could continue to the next page in the pre-test, main task, and post-test only after a few seconds. This design choice followed the protocol of \cite{spitzer2024effect} and was to ensure that participants focused on the actual task. 
In the final part (5), participants answered a questionnaire. The questionnaire asked them to describe and explain each architectural style. With this, we were able to analyze whether participants had learned the correct distinctions between the architectural styles and were able to conclude based on their understanding (e.g., their reasoning). On top of that, they had to rate several variables on seven-point Likert scales to measure for confounding factors. These included the AI usefulness, their experience with AI, their cognitive load, and their trust in AI. All variables are presented in \Cref{sec: appendix} in \Cref{tab: items}. Throughout the questionnaire, we implemented a second attention check to ensure only valid results, as suggested by \citet{abbey2017attention}. In this attention check, we asked participants to select \textit{''Likely``} for an item of AI usability (\textit{''As an attention check, please choose "Likely" for this statement``}). The questionnaire ended with demographic questions on participants' age, gender, education and employment.

\subsection{AI Development}
 
The AI used in the study was a LLM that provided participants with a classification (AI advice) and an explanation for its reasoning, depending on the treatment they were assigned to. As LLM, we used Open AI's GPT-4o model (model version 2024-05-13) through an Azure Open AI Studio instance. In the pre-phase of the study, we tested multiple prompt strategies in a workshop with three authors. The final prompts that were used are shown in \Cref{sec: appendix} in \Cref{tab: prompts}. The LLM was prompted such that the correct explanation treatment provided an explanation that corresponded to its prediction, while the explanation did not match the prediction in the incorrect explanation treatment. 

The definition of an incorrect explanation follows \citet{morrison2024impact} and \citet{cabitza2024explanations} in that the AI does not provide correct reasoning for its predicted class (not coherent with the AI advice), independent of the ground truth of the image. For instance, if the AI made a correct prediction for an Art Nouveau building, but the explanation does not conform to this class, it is defined as incorrect. The definition further follows \citet{cabitza2024explanations} in that the incorrect explanations are pertinent as they emphasize features that are not relevant for a given predicted architecture style. For instance, if the incorrect explanation highlights features of an Art Nouveau building that are actually characteristic of Art Deco. To avoid information overload \cite{arnold2023dealing}, we designed the explanations to be no longer than three sentences. As outlined in \Cref{sec:theoretical-development}, we analyze the impact of incorrect explanations for cases where the AI classification is correct.

\subsection{Recruitment}

We recruited 186 participants from the United States through the platform Prolific.co and ran the study on August 12, 2024. Previous research indicates that this platform is a reliable source of research data \citep{peer2017beyond, PALAN201822}. Several screening mechanisms were implemented through the Prolific platform. With the filters, we targeted individuals who were fluent in the English language and had shown high quality in previous studies (100\% completion rate). Our recruitment strategy was designed to not focus on participants with specific backgrounds, but admit participants without any further restrictions to be able to generalize our findings. 
Participants who met the stated criteria and completed the study's requirements received a base payment of $2.25$\pounds. Additionally, we implemented an incentive structure: participants are incentivized to conduct the task correctly by providing a bonus for each correctly classified image. This should ensure that participants paid attention during the task and did not provide random answers. The bonus was 4 pennies for each correct answer and led to a potential maximum payment of $3.21$\pounds. As stated in our pre-registration, we excluded participants who did not finish the main task on time (within 30 minutes) or did not finish the entire study. We also excluded participants with obvious misbehavior (e.g., clicking through the cases and always providing the same answer). Additionally, we computed the overall mean and standard deviation across all treatments and winsorized at 2.5 SD above/below the mean. Applying this exclusion strategy, we ended up with 160 participants equally assigned to the four treatments (40 participants for each treatment).

\subsection{Metrics}

Similar to previous work (e.g.,  \cite{hemmer2023human,schoeffer2024explanations}), we assessed participants' task performance in classifying the architectural styles in pre-test, main task, and post-test. Aligning with prior research \citep{mccormick1997conceptual, nahdi2020conceptual}, we use the task performance to infer participants' procedural knowledge. We measure the human-AI team performance in the main task. As metrics for the task performance, we used accuracy and measured the ratio of correctly classified images over all images. Participants had to select one of the three different architectural styles for each image by selecting from a drop-down menu on each page of the task. This means that a random guess corresponded to 33.3\% of performance. Aligning with \citet{schoeffer2024explanations}, we also measured the correct adherence and detrimental overrides in the main task of the study.

We assessed participants' reasoning ability for the three architectural styles through open-ended questions by asking them to describe and explain each style, thereby following the procedure of \citet{chi1989self}. We assessed participants' reasoning abilities in terms of correctness by comparing them to the correct characteristics and features for each architectural style. We followed the procedure of \citet{huang2024chatgpt} and used LLMs to evaluate the correctness of participants' answers and computed reasoning scores for participants: zero represents a completely incorrect or irrelevant answer, and five represents a completely correct and comprehensive answer. The prompt used for this analysis is in \Cref{tab: prompts} in \Cref{sec: appendix}. By doing so, we were able to assess whether incorrect explanations impaired participants' reasoning.

Finally, we established several control variables to investigate the potential underlying factors that might influence AI-assisted decision-making. In particular, we controlled for participants' cognitive load as previous research suggests that the information in explanations displayed to humans can affect their decision-making behavior \cite{abdul2020cogam, hudon2021explainable, herm2023impact, spitzer2024effect}. We measured participants' cognitive load on a seven-point Likert scale by having them rate five validated items previously developed by \citet{hart1986nasa} and that were already applied in the field \cite{senoner2024explainable}. In addition, we assessed participants' AI trust, AI usefulness, and experience with AI by using items proven in previous research \cite{senoner2024explainable}, all of them also rated on a seven-point Likert scale. All items are shown in \Cref{tab: items} in \Cref{sec: appendix}.

%% file: 05-results.tex
\section{Results}
\label{sec: results}

It took participants, on average, 14 minutes and 54 seconds to complete the study. Overall, 160 participants passed the attention check and finished the study according to the study protocol. Of these 160 participants, 79 were male, 77 were female, and four identified as diverse. 
To address the research questions, we first conduct several statistical analyses to answer \textbf{RQ: \ref{rq1}} in subsection \ref{sec: results-proceduralknowledge}. Subsequently, we address \textbf{RQ: \ref{rq2}} and qualitatively assess the open-ended questionnaires in subsection \ref{sec: results-reasoning}. In the final subsection \ref{sec: results-team}, we evaluate the impact of incorrect explanations on the human-AI team performance (\textbf{RQ: \ref{rq3}}).
%In this section, we provide not-corrected and corrected results of the statistical tests. As we only run the statistical test for a maximum number of three hypotheses for each analysis, the correction is not necessarily required, as highlighted by research from the statistical field. Yet, the corrected numbers provide a more conservative view, minimizing type 2-error in the tests. (ADD additional explanation by GPT). We draw our conclusions based on the corrected values, thereby providing robust results.

\subsection{RQ1: Impact of Incorrect Explanations on Procedural Knowledge}
\label{sec: results-proceduralknowledge}

To establish a baseline and ensure that all treatments began on an equal performance level in classifying the architectural styles, we conduct a one-way ANOVA on the pre-test performance scores across the four treatments ($F = 0.80, p = .498$). These results fail to reject the null hypothesis, indicating no significant differences in pre-test performance among the four treatments. This finding suggests that participants across all treatments start with comparable levels of procedural knowledge to classify the architectural styles prior to the main task. This allows for a thorough interpretation of any differences observed in the post-test results, as they can be more readily attributed to the treatment effects rather than pre-existing differences.

\begin{figure}[htbp!]

\centering
\begin{subfigure}{0.48\textwidth}

  \centering
  \includegraphics[width=\textwidth]{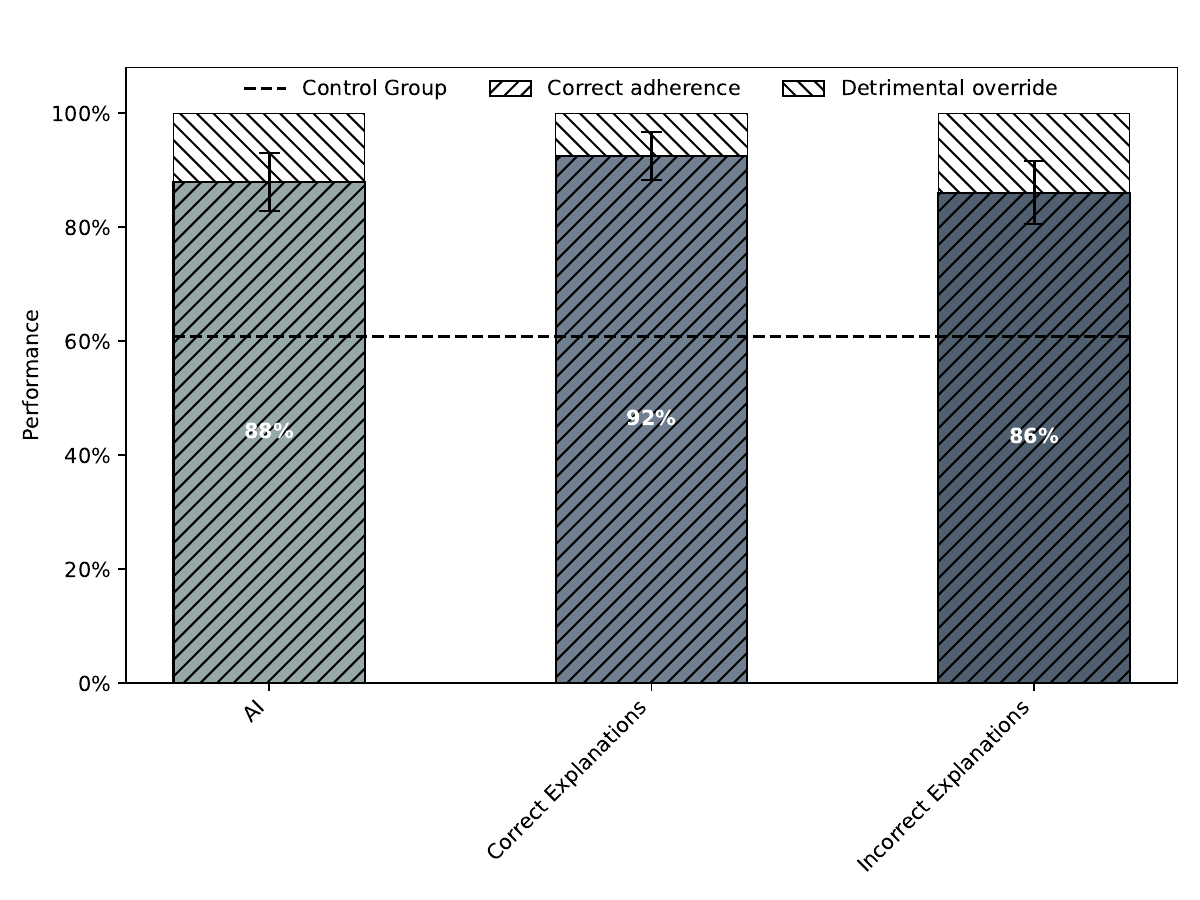}
    \captionsetup{justification=centering}

  \caption{The main task performances across treatments.}
  \label{maintaskperformance}
\end{subfigure}
\begin{subfigure}{0.48\textwidth}
  \centering
  \includegraphics[width=\textwidth]{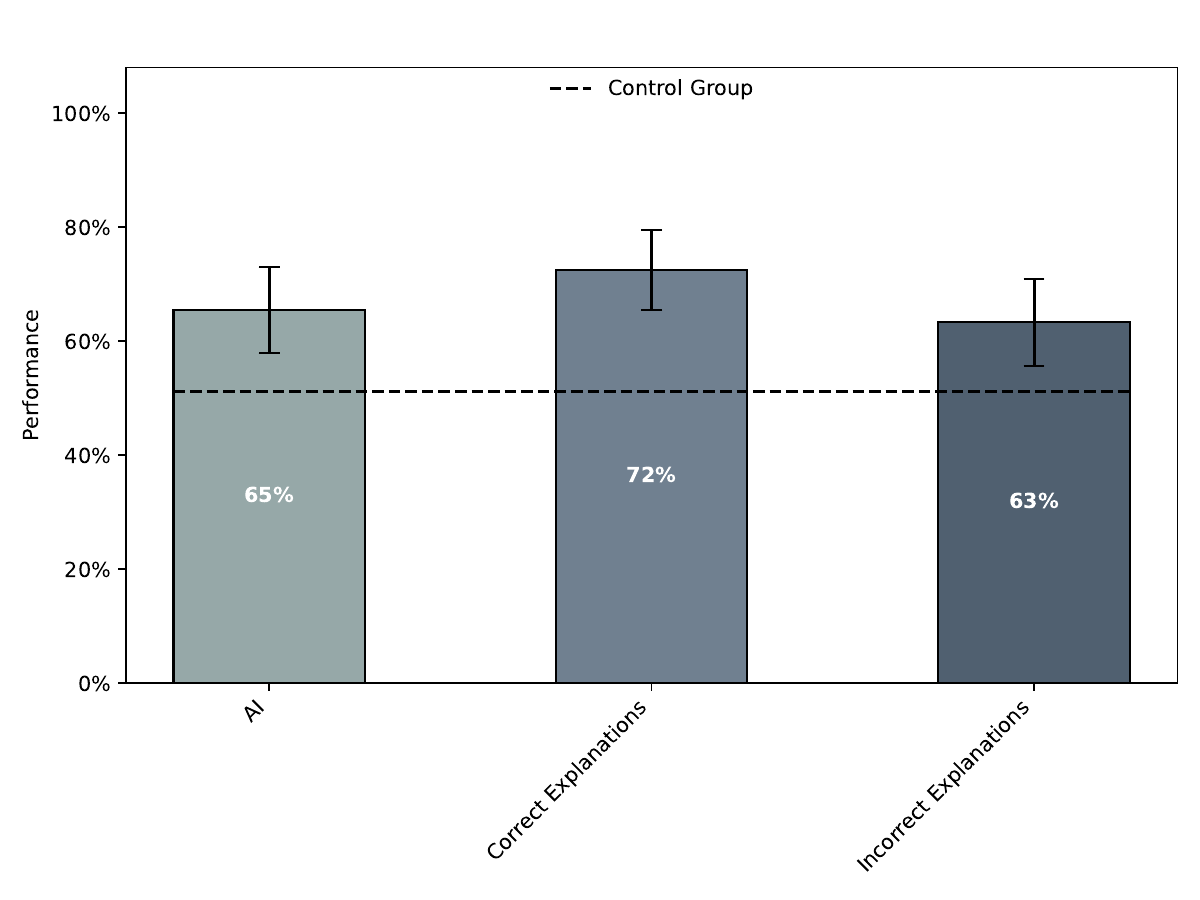}
    \captionsetup{justification=centering}

  \caption{The post-test performances across treatments.}
  \label{posttestperformance}
  \centering
\end{subfigure}

    \centering
\captionsetup{justification=raggedright,singlelinecheck=false}
\caption{The subfigures present the performances across treatments in the main task and post-test of the study.}
\Description{The figure shows two subfigures. In subfigure A, the performance of participants across treatments is shown for the main task. In subfigure B, for the post-test.}
\label{performances}

\end{figure}

Overall, the control group maintains the lowest performance at around $51.25\%$, while the correct explanation treatment continues to lead with approximately $72.50\%$ accuracy. The AI classification and incorrect explanation treatments show similar post-test performances of about $65.42\%$ and $63.33\%$, respectively. 
%We can also see that with the different performance levels in each treatment, the correct adherence and the detrimental overrides change. \textbf{Especially in the incorrect explanation treatment, the detrimental overrides are the highest during post-test, indicating that participants' procedural knowledge in this treatment is impaired the most.}

To assess the significance of differences in post-test performance across treatments, we compare performance levels between treatments. A one-way ANOVA yields evidence of treatment effects ($F = 5.86, p = .001$), indicating that the type of AI support in the main task influences participant performance in the post-test. To further examine these differences, we conduct pairwise comparisons using one-sided t-tests, assuming participants in the incorrect explanation treatment exhibit lower performance compared to participants in the other treatment. We report the corrected p-values according to the Holm-Bonferroni correction \cite{holm1979simple} (see \Cref{tab: posttestperf} in \Cref{sec: appendix}). 
There is no significant difference between the post-test performances in the incorrect explanation treatment and the control group ($t = 2.241, p = .986$). Comparison with the AI treatment also shows no significant difference ($t = -.371, p = .356$). Notably, the comparison between incorrect and correct explanations yields a significant difference ($t = -1.742, p = .085$), with incorrect explanations leading to inferior performance. With these findings, we find support for hypothesis \ref{hyp1}. \textbf{These results demonstrate that the correctness of explanations plays an important role in AI-assisted decision-making, with incorrect explanations decreasing the task performance of humans in the post-test the most}. Correct explanations provide a performance advantage over incorrect explanations, underscoring the importance of explanation accuracy in AI-assisted tasks. This is an interesting finding: we compare the post-test performance levels between the treatments in a one-sided manner and assume that the incorrect explanation of treatment leads to a performance level lower than that of the other treatments. While we can see that participants provided with incorrect explanations perform worse, correct explanations lead to the highest post-test performance. Contrarily, compared to having no explanation or no AI advice at all, this effect diminishes, potentially showing that incorrect explanations do not lower the post-test performance as much as when correct explanations are provided.

To identify the potential reasons behind these findings, we further analyze which control variables influence the post-test performance.
We run a regression analysis with the post-test performance as the dependent variable and model the type of AI support as the independent variable. We define cognitive load, AI trust, AI knowledge, and AI usefulness as confounding factors. 
The regression analysis examining the impact of different treatments on post-test performance reveals that the type of AI support significantly influences participants' performance (see \Cref{tab: regressionposttest} in \Cref{sec: appendix}). The model accounts for $12.7\%$ of the variance in post-test performance ($R² = .127, p = .004$). Specifically, participants in the correct explanation treatment exhibit the greatest improvement in post-test performance, with a significant positive effect ($coef = 0.198, p = .000$), followed by those in the AI treatment ($coef = 0.125, p = .020$) and the incorrect explanation treatment ($coef = 0.117, p = .025$). The data supports the findings of the statistical tests. \textbf{Interestingly, cognitive load is negatively associated with post-test performance ($coef = -.197, p = .097$), suggesting that higher cognitive load impedes procedural knowledge.} We further analyze this finding by comparing the cognitive load for each treatment. We can see that cognitive load is highest for the control group ($35.42\%$) and the incorrect explanation treatment ($33.58\%$) compared to the other treatments (AI treatment = $28.75\%$, correct explanations treatment = $28.33\%$) illustrating that the decrease in procedural knowledge stems from the increased cognitive load. Contrarily, trust, AI knowledge, and AI usefulness do not impact the post-test performance.
%The results indicate that while the incorrect explanation treatment improves the post-test performance compared to the control group, it was less effective than the correct explanations and AI-only support. \textbf{Specifically, participants in the incorrect explanation treatment achieve a lower post-test performance than those in the AI treatment, indicating that incorrect explanations lead to a lower level of procedural knowledge.} 

\begin{figure}[htbp!]

  \centering
  \includegraphics[width=.75\textwidth]{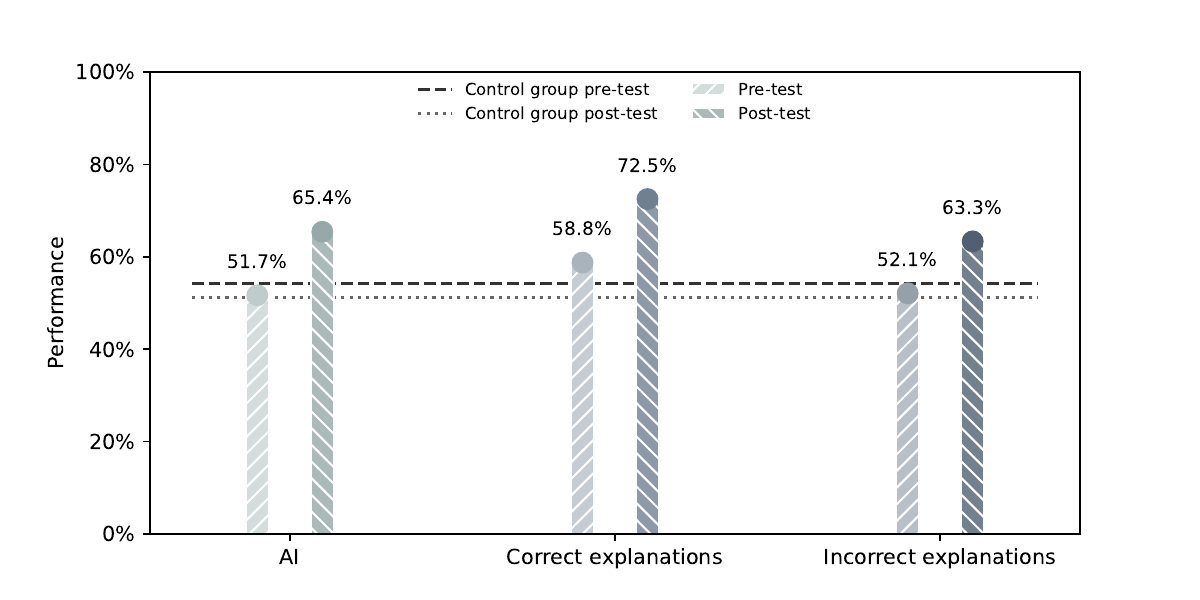}
    \captionsetup{justification=centering}

    \centering
\captionsetup{justification=raggedright,singlelinecheck=false}
\caption{The procedural knowledge approximated by task performance in pre-test and post-test across the different treatments. The control group is highlighted as line plots.}
\Description{The figure shows the difference in performance between the pre-test and post-test for the different treatments as bar plots. Additionally, the performance of the control group in pre-test and post-test is shown as line plots.}
\label{fig: learning_gain}

\end{figure}

Additionally, we analyze how the procedural knowledge develops from pre-test to post-test. We plot this difference in \Cref{fig: learning_gain}. 
The figure illustrates whether procedural knowledge develops or degrades across different decision-making stages in the different treatments. Interestingly, the data reveals that the incorrect explanation treatment still leads to procedural knowledge development. \textbf{Conversely, the AI treatment and correct explanation treatment show the highest procedural knowledge development, exceeding the development in the incorrect explanation treatment. }

%This suggests that while incorrect explanations may temporarily enhance task performance, they ultimately hinder the development of procedural knowledge, leading to poorer long-term retention.

%This supports the previous findings that correct explanations significantly enhance procedural knowledge and promote better long-term retention compared to incorrect explanations.

% Furthermore, we run an ANOVA which reveals significant differences in procedural knowledge gains across treatments ($F = 4.04, p = .008$). To identify the differences between the treatments we subsequently conduct a Tukey HSD Post-Hoc Test, which accounts for the multiple comparisons, and, thus provides corrected p_values. Notably, ...

Furthermore, we conduct an ANOVA and Tukey HSD Post-Hoc Test which reveal significant differences in procedural knowledge gains across treatments ($F = 4.04, p = .008$). Notably, the incorrect explanation treatment does not significantly differ from the AI treatment or the correct explanation treatment ($p = 0.9707$ for both comparisons), but it shows a trend toward higher procedural knowledge gains compared to the control group ($p = .061$). The procedural knowledge development is significantly higher than in the control group ($p = .019$). This suggests that while incorrect explanations may increase procedural knowledge in collaboration with an AI, the correctness of explanations reduces the effect compared to correct explanations.
%\textbf{The findings imply that while AI support is generally beneficial, the accuracy of explanations plays a critical role in ensuring these benefits extend to longer-term knowledge retention.}

\subsection{RQ2: Impact of Incorrect Explanations on Reasoning}
\label{sec: results-reasoning}

To analyze participants' reasoning capabilities as they transition from knowing "how" to understanding "why", we assess the answers they provided in the open-text questionnaire. The primary focus is on evaluating how the accuracy of AI-generated explanations influences the participants' reasoning abilities to understand how incorrect explanations impact their knowledge. We provide some randomly selected answers to these open-ended questions in the \Cref{sec: appendix} in \Cref{tab: reasoninganswers}. All participants' results are provided in the supplementary materials. 

The scores are then analyzed using independent t-tests to compare the performance between treatments, and the Holm-Bonferroni method is used to correct them.

In the incorrect explanation treatment ($score = 54.58\%$), participants score lower on average than those in the control group ($score = 57.20\%$). However, the difference is not significant ($t = -.706, p = .241$). 
When comparing the incorrect explanation treatment to the AI treatment($score = 62.38\%$), participants exposed to incorrect explanations perform significantly worse ($t = -2.203, p = .046$). 
The comparison between the incorrect explanation treatment and the correct explanation treatment ($score = 60.00\%$) shows similar results ($t = -1.645, p = .104$). While not significant at alpha = 0.1, \textbf{we see a tendency for incorrect explanations to decrease reasoning abilities compared to correct explanations or only AI advice.}
Participants exposed to incorrect explanations decrease their reasoning ability. With this decrease in the incorrect explanation treatment compared to the correct explanations and AI advice only treatment, there is no significant difference compared to the control group. This means that receiving incorrect explanations has no difference in its impact on participants’ understanding than no AI support. This could mean that the incorrect explanations might still be helpful in certain scenarios to participants to identify actual correct characteristics of the architectural styles. 

\begin{figure}[htbp!]

\includegraphics[width=.5\linewidth]{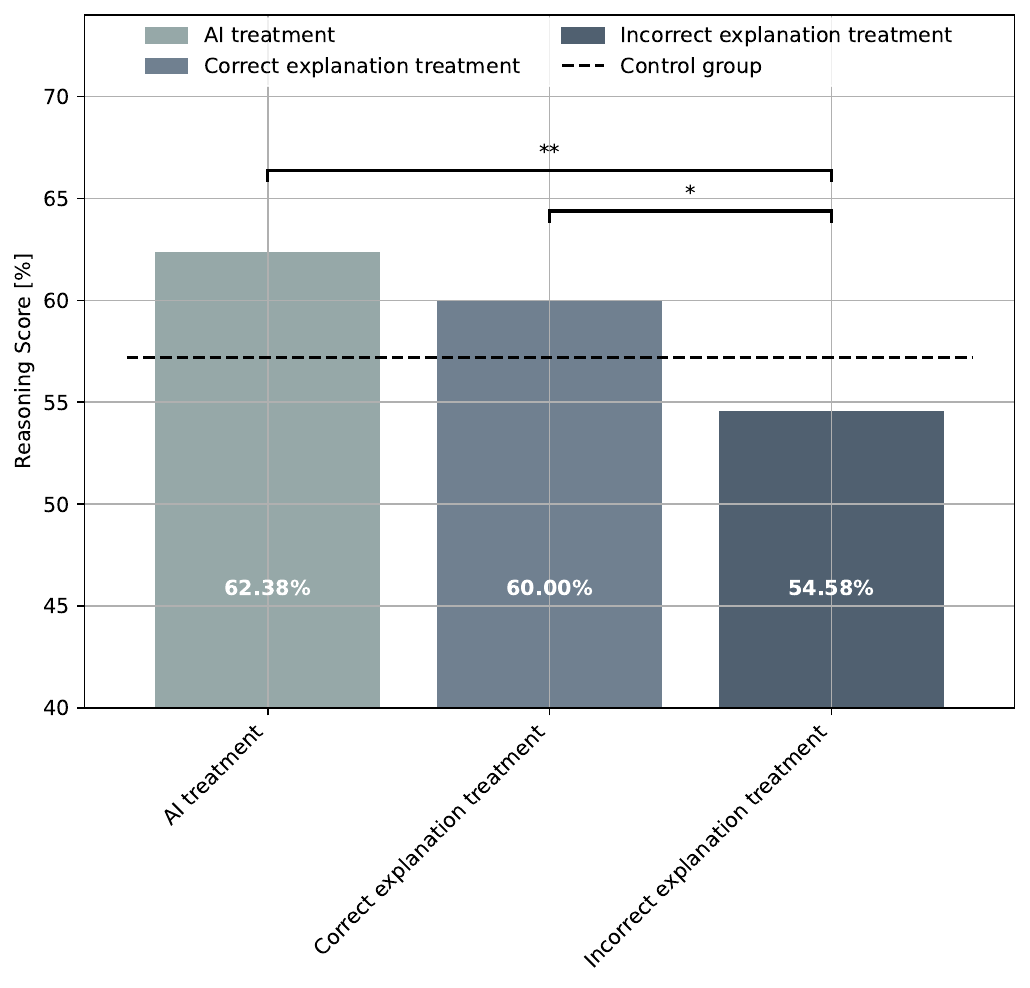}
\captionsetup{justification=raggedright,singlelinecheck=false}
\caption{The reasoning scores for participants of the different treatments. (\textit{*} \textit{p < .1}; \textit{**} \textit{p < .05};  \textit{***} \textit{p < .01})}
\Description{The figure shows bar plots for the different treatments regarding the reasoning scores calculated. Additionally, a dotted line is shown that represents the control group.}
\label{fig: reasoning}

\end{figure}

Additionally, we analyze further control variables that might influence participants’ reasoning capabilities. To do so, we run a regression analysis, similar to \Cref{sec: results-proceduralknowledge}, using cognitive load, trust, AI knowledge, and AI usefulness as control variables. We find that AI usefulness has a direct effect on the reasoning capabilities ($coef=.7443$, $p=.068$). We report the results in \Cref{sec: appendix} in \Cref{tab: regressionreasoning}. Thus, when participants perceive the AI as useful, they also build a better understanding of the underlying task domain. This is an interesting finding that we further discuss in \Cref{sec: discussion}.

\subsection{RQ3: Impact of Incorrect Explanations on Human-AI Team Performance}
\label{sec: results-team}

In this subsection, we analyze the data of the study to derive insights into how incorrect explanations affect the human-AI team performance. 

The main task performance results reveal substantial differences across treatments (see \Cref{performances}). The control group demonstrates the lowest performance at $60.83\%$. In contrast, all AI-assisted treatments show higher performance levels. The AI treatment achieves $87.92\%$ accuracy, while the correct explanation treatment performs best at roughly $92.50\%$. Interestingly, the incorrect explanation treatment still outperforms the control group, reaching about $86.04\%$ accuracy. These patterns persist, albeit with reduced magnitudes, in the post-test performance (see \Cref{sec: results-proceduralknowledge}. We can also see that with the different performance levels in each treatment, the correct adherence and the detrimental overrides change. \textbf{Especially in the incorrect explanation treatment, the detrimental overrides are the highest, indicating that participants overrely on the AI.}
\textbf{These results suggest that AI support, regardless of the explanation accuracy, enhances performance during the main task, with incorrect explanations leading to the lowest gain.} 

To assess the significance of differences in main task performance across treatments, we conduct a one-way ANOVA. The results ($F = 27.80, p = .000$) indicate significant differences among the treatments. To assess the impact of incorrect explanations with the other treatments on main task performance, we conduct pairwise comparisons between the incorrect explanation treatment and other treatments using one-sided t-tests, assuming that participants in the incorrect explanation treatment exhibit lower performance compared to participants in the other treatments (see hypothesis \ref{hyp3}). We also correct the tests using the Holm-Bonferroni method and report these p-values (see \Cref{tab: maintaskperf} in \Cref{sec: appendix}). 
The comparison between the incorrect explanation treatment and the control group shows no significant difference in performance ($t = 5.637, p = 1.000$), indicating that the performance for incorrect explanations is not below the control group's. Similarly, the comparison with the AI treatment yields no significant difference ($t = -.546, p = .293$). However, when comparing the incorrect explanation treatment with the correct explanation treatment, there is a significant difference in main task performance ($t = -2.359, p = .021$), \textbf{indicating that the incorrect explanations lead to a lower performance compared to the correct explanations.} Thus, we find support for hypothesis \ref{hyp3}.

Analog to the post-test performance, we run a regression analysis to reveal the effect of confounding factors on the main task performance (see \Cref{tab: regressionmain} in \Cref{sec: appendix}). The model investigating the main task performance explains $36.7\%$ of the variance ($R² = 0.367, p = .000$), showing that participants in AI-supported treatments outperform those in the control group.
Participants in the correct explanation treatment demonstrate the most substantial improvement over the control group, with a significant positive effect ($coef = .311, p = .000$). This is closely followed by the AI treatment ($coef = 0.259, p = .000$) and the incorrect explanation treatment ($coef = 0.250, p = .000$). 
Furthermore, AI trust shows a significant negative association with performance ($coef = -.237, p = .078$), indicating a possible nuanced effect of trust on how participants interact with the AI. We take further steps and analyze AI trust between the different treatments. AI trust is highest for the correct explanation treatment ($48.75\%$), followed by the incorrect explanation treatment ($47.08\%$), the control group ($46.32\%$), and the AI treatment ($44.72\%$). This finding shows that explanations increase trust independently of their correctness. Interestingly, as the score is higher in the control group than in the AI treatment, only providing AI advice does not seem to benefit participants' trust in the AI.

The findings suggest that while incorrect explanations can aid immediate task performance, they may hinder the retention of procedural knowledge (as the post-test performance is below the main-task performance). 
We use a repeated measures approach and run a mixed-effects model to analyze how procedural knowledge is affected by first providing and then removing AI support. We define performance as the dependent variable, the type of AI support as the independent variable, the different stages---with AI support in the main task and without AI support in the post-test---as the mediating factor, cognitive load, AI trust, AI knowledge and AI usefulness as control factors and participant ID as a random factor.

\begin{table}[htbp!]
\caption{Mixed Effects Model Analysis on Performance}
\Description{The table shows the results of the mixed effects model with performance as the dependent variable. In the columns, the coefficient and the standard error are presented. As significance levels, .1, .05, and 0.01 are defined.}
\begin{threeparttable}
\begin{tabular}{m{7cm} R{1.5cm} R{1.5cm}} \hline
\multicolumn{1}{m{7cm}}{Dependent variable} & \multicolumn{2}{c}{Performance} \\
\cmidrule{2-3}
& Coeff & SE \\
\hline \hline
Intercept                                         & 0.695***   & 0.084  \\
AI support: & & \\
\textit{- control group (baseline)} && \\
\textit{- AI treatment}                             & 0.256***   & 0.046  \\
\textit{- correct explanation treatment}             & 0.306***   & 0.047  \\
\textit{-incorrect explanation treatment}          & 0.249***   & 0.045  \\
decision-making stage                                        & -0.096***   & 0.035  \\
AI treatment :decision-making stage                      & -0.129***   & 0.050  \\
correct explanation treatment:decision-making stage        & -0.104**    & 0.050  \\
incorrect explanation treatment:decision-making stage      & -0.131***  & 0.050  \\
cognitive\_load                                   & -0.110     & 0.086  \\
AI trust                                        & -0.187     & 0.132  \\
AI knowledge                                     & -0.011     & 0.060  \\
AI usefulness                                      & 0.063      & 0.095  \\
participant\_ID                               & 0.016      & 0.029  \\ \hline
Log-Likelihood                                    & \multicolumn{2}{c}{49.9483} \\
Scale                                             & \multicolumn{2}{c}{0.0248} \\
\hline
\end{tabular}
    \begin{tablenotes}
        \item[1] Note: \textit{*} \textit{p < .1}; \textit{**} \textit{p < .05};  \textit{***} \textit{p < .01}\newline\newline
    \end{tablenotes}
\end{threeparttable}
\end{table}

\begin{figure}[htbp!]

\centering
\begin{subfigure}{0.31\textwidth}

  \centering
  \includegraphics[width=\textwidth]{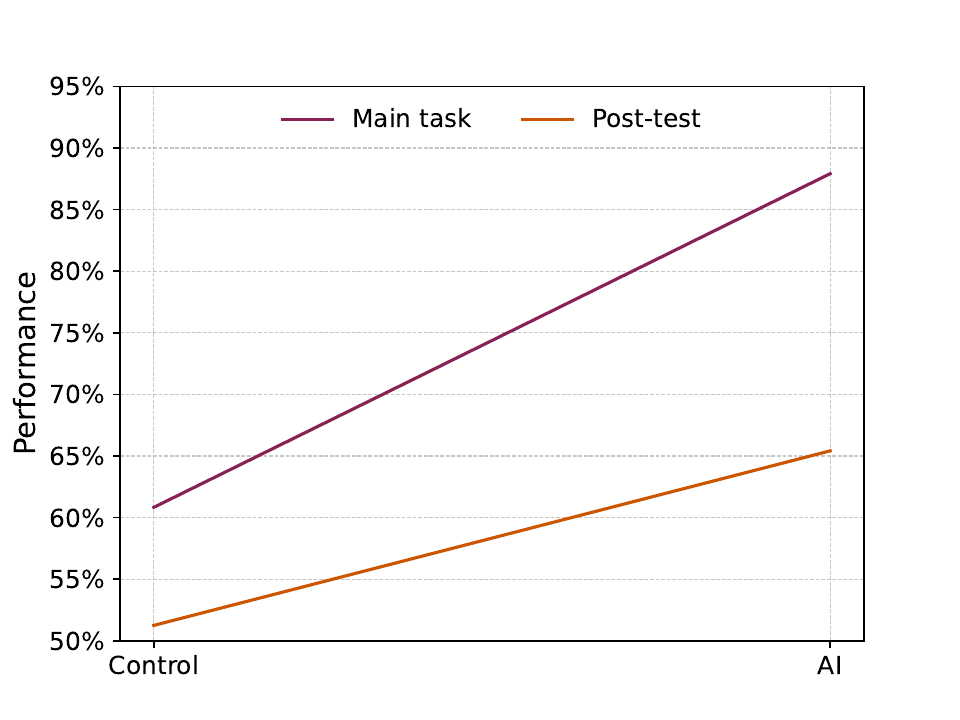}
    \captionsetup{justification=centering}
    \label{subfig: AI}
  \caption{The interaction effect on the AI treatment.}
\end{subfigure}
\begin{subfigure}{0.31\textwidth}
  \centering
  \includegraphics[width=\textwidth]{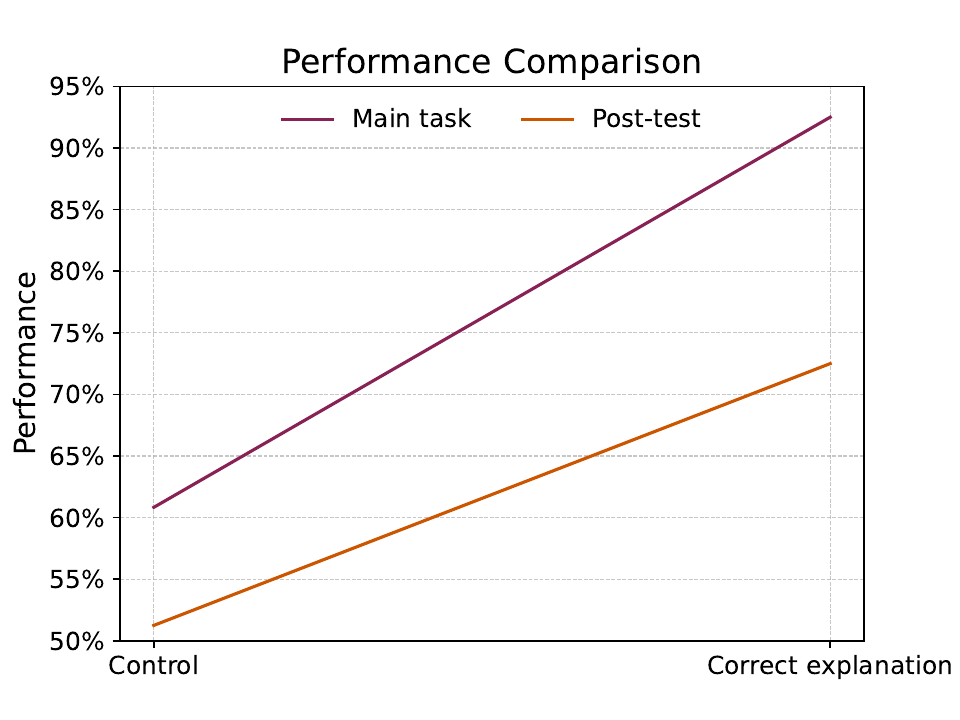}
    \captionsetup{justification=centering}
    \label{subfig: correct}
  \caption{The interaction effect on the correct explanation treatment.}
  \centering
\end{subfigure}
\begin{subfigure}{0.31\textwidth}
  \centering
  \includegraphics[width=\textwidth]{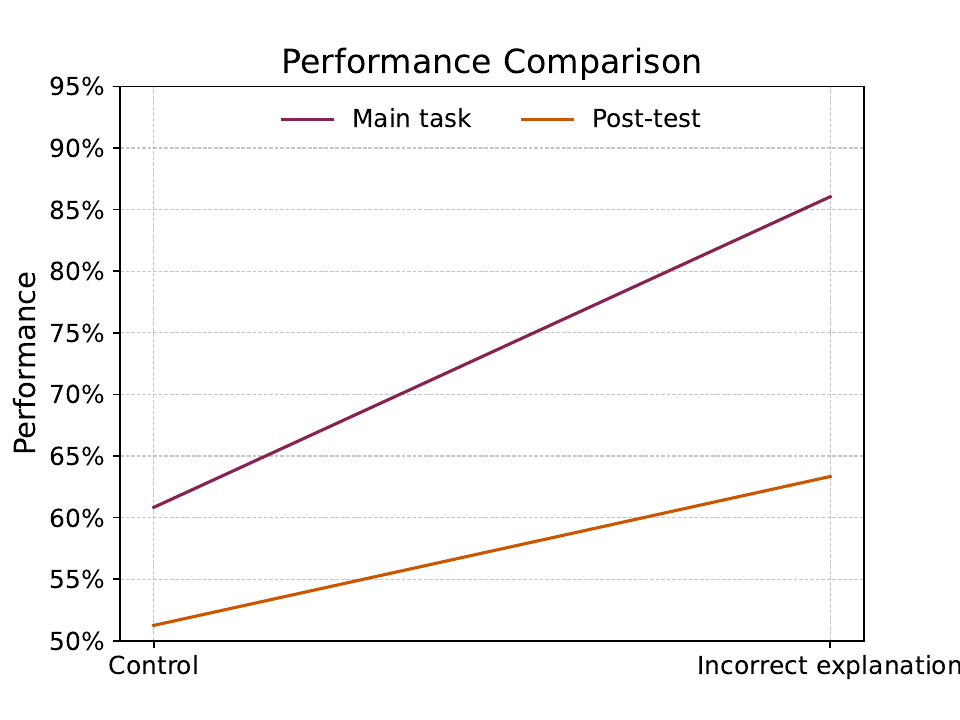}
    \captionsetup{justification=centering}
    \label{subfig: incorrect}
  \caption{The interaction effect on the incorrect explanation treatment.}
  \centering
\end{subfigure}

    \centering
\captionsetup{justification=raggedright,singlelinecheck=false}
\caption{The sub-figures present the interaction effects of the transition from the main task to the post-test on the relationship of the type of AI support on performance.}
\Description{The figure shows three sub-figures. In sub-figure A, the interaction effect of the transition from main task to post-test on the AI treatment, in figure B on the correct explanation treatment and in figure C on the incorrect explanation treatment.}
\label{fig: interactions}

\end{figure}

Participants in all AI treatments show significantly higher performance during the main task compared to the control group, with the correct explanation treatment leading to the highest performance ($coef = .306, p = .000$), followed by the AI treatment ($coef = .256, p = .000$), and the incorrect explanations ($coef = .249, p = .000$). However, when transitioning to the post-test, where AI support is removed, all AI-assisted treatments experience a significant decline in performance. The decrease is most pronounced for the incorrect explanation treatment ($coef = -.131, p = .008$), \textbf{indicating that while AI support with incorrect explanations initially boosts the performance during the main task, it negatively impacts the retention and development of procedural knowledge when the AI support is removed}. The interaction effects in \Cref{fig: interactions} support this. The sub-figures illustrate that incorrect explanations can undermine procedural knowledge development, as evidenced by the greater performance drop in the post-test for participants who receive incorrect explanations (see \Cref{subfig: incorrect}) compared to those who receive correct explanations (see \Cref{subfig: correct}) or AI advice only (see \Cref{subfig: AI}) (lower slope in the lines for incorrect explanations). 

%% file: 06-discussion-limitations.tex
\section{Discussion}
\label{sec: discussion}

\subsection{Summary of Findings}

With the rise of AI in decision-making domains, it is crucial to understand how the interaction with AI affects decision-makers. Prior research so far has either focused on the implications of incorrect AI advice (e.g., \cite{schemmer2023appropriate, bansal2021does}) or explored how the correctness of explanations affects trust and reliance on AI (e.g., \cite{morrison2024impact, cabitza2024explanations}). This work investigates how the correctness of AI explanations impacts humans through collaboration with an AI. In a classification task, participants were assigned to one of four treatments: no AI support, AI advice only, AI advice with correct explanations, and AI advice with incorrect explanations. We measured the task performance before, during, and post-collaboration to derive humans' procedural knowledge, reasoning, and the human-AI team performance.
The results show that all AI-assisted treatments led to significantly higher human-AI team performance during the main task compared to the control group, demonstrating the merits of AI support. However, the correctness of the explanations played a crucial role:
Participants who had received correct explanations exhibited the highest main task performance, followed by those who received only AI advice without explanations, dominating those who had received incorrect explanations.
Importantly, the findings reveal that the accuracy of the explanations has a lasting impact on procedural knowledge development. While incorrect explanations temporarily increase performance during the main task, the procedural knowledge was impaired post-collaboration, ultimately hindering knowledge retention compared to correct explanations or no explanations. Moreover, these results must be considered with caution. Even though there was an increase in procedural knowledge, participants' reasoning capabilities in the incorrect explanation treatment were below the ones from the control group.
This indicates that incorrect explanations can undermine the durable benefits of human-AI collaboration despite initial performance improvements. 
These results contribute to the HCI literature by underscoring the nuanced effects of explanation accuracy on human-AI collaboration. The findings emphasize the importance of designing AI-based support that provides accurate explanations to not only enhance human decision-making but also maintain and improve humans' procedural knowledge and reasoning in the long run.
Overall, the study's key takeaway is that, while AI support can generally improve performance, the correctness of the explanations provided by the AI is a crucial determinant of human-AI collaboration, influencing humans' ability to draw conclusions and perform the task autonomously post-collaboration with an AI. These insights can inform the design of more effective and human-centric AI-based decision support systems.

\subsection{Implications}

%------intro-------
This work makes several contributions to the field of HCI by deepening the understanding of how the correctness of AI explanations influences humans' procedural knowledge and reasoning ability, as well as the human-AI team performance. Although previous research has often focused on the benefits of AI assistance in enhancing decision-making \cite{schemmer2023towards}, this study offers a more nuanced perspective, highlighting the critical role of explanation correctness on humans' knowledge and understanding. It complements the large corpus of HCI literature on XAI (i.e., \cite{morrison2024impact, schoeffer2024explanations, speith2022review}) by explicitly investigating the influence of incorrect explanations for scenarios when the AI advice is correct and identifies the negative repercussions for humans.

%------misinformation effect-------
\textit{The misinformation effect of explanations.} The study highlights a crucial risk associated with AI explanations: the potential \textbf{misinformation effect} to impair procedural knowledge and reasoning through incorrect explanations. The misinformation effect, extensively studied by \citet{loftus1978semantic, loftus1974reconstruction}, describes how exposure to incorrect information can distort an individual's memory of an event. This phenomenon extends to AI explanations, where incorrect explanations can similarly distort humans' understanding. The integration of incorrect information through explanations into existing knowledge structures, as discussed by \citet{ayers1998theoretical}, can lead to significant changes in both procedural knowledge and reasoning.
Incorrect explanations, while potentially unharmful in the short term, can create an illusion of understanding, resulting in poorer performance in subsequent tasks without AI support. This phenomenon is particularly concerning in high-stakes environments such as healthcare or legal decision-making, where the quality of decision-making has far-reaching consequences \cite{topol2019high}. For organizations, this implies that the long-term efficacy of AI hinges not only on their immediate performance but also on their ability to foster accurate knowledge. Incorrect explanations can lead to a misalignment between the AI’s recommendations and humans’ understanding, which may diminish their overall effectiveness \cite{morrison2024impact}. Yet, it also has the potential to impair humans post-collaboration with an AI. Therefore, organizations must prioritize the development of AI that provides correct and transparent explanations to ensure sustained, high-quality decision-making and prevent detrimental impacts on organizational knowledge and performance.

%------impact on task performance-------
\textit{Taking a human-centric perspective.} In our study, we could see that the human-AI team performance improved over the participants' initial task performance. This pattern, also observed by prior research \cite{inkpen2023advancing, hemmer2024complementarity}, showcases the potential of humans collaborating with AI. Even though the scenario in which the human-AI team performance exceeds the performance of human or AI alone---also referred to as complementary team performance \cite{bansal2019beyond}---could not be reached (due to the study design choices this was not feasible as the AI advice was always correct), our findings showcase the two sides of explanations: while correct explanations improve the human-AI team performance compared to only receiving AI advice (bright side) and demonstrate the merits of XAI, incorrect explanations decrease the human-AI team performance compared to only receiving AI advice (dark side) and outline the pitfalls of XAI. Thus, it is important to implement mechanisms that allow humans to verify the correctness of explanations, for instance, through reflection mechanisms \cite{ehsan2024explainability}. Furthermore, we could also identify AI trust as a influencing factor during the main task and cognitive load during the post-test. While these findings align with prior research \cite{ueno2022trust, buccinca2021trust, westphal2023decision}, they also emphasize the important role of \textbf{individuals' characteristics and traits} in human-AI collaboration. Therefore, research has to anticipate these factors in the design of robust and safe explanations.

%------The role of the AI-------
\textit{The role of AI.} While the focus of our research has mainly been on the humans in our AI-assisted decision-making study, the AI can take an important role in \textbf{minimizing the risk of negative repercussions}. In our study, these negative repercussions were expressed by humans' decreasing procedural knowledge and reasoning capabilities post-collaboration. Although this decline was seen in every AI-supported treatment, incorrect explanations led to the greatest decline, with reasoning ability falling below that of the control group. Recent research in HCI should, therefore, focus on developing methods in which the AI itself can warn human collaborators of a potentially incorrect explanation, for instance, by using uncertainty scores \cite{huang2023look, prabhudesai2023understanding} or cognitive forcing functions \cite{buccinca2021trust}. 

%------Anchoring on Explanations-------
\textit{Anchoring on explanations. }We also take a critical view on the phenomenon prior research has identified in collaboration scenarios between humans and AI in which the AI provides explanations: the anchoring effect on explanations \cite{wang2019designing, bansal2021does}. The effect occurs when human decision-makers fixate on the explanation and form an incorrect understanding. In our study, participants in the incorrect explanation treatment demonstrated this behavior by achieving lower procedural knowledge on the post-test compared to the other AI-supported treatments and also by showing lower reasoning capabilities than participants in the control group. As a result, they made \textbf{the right decisions for the wrong reasons}, a condition often referred to as the Clever Hans Effect. \cite{schramowski2020making, kauffmann2020clever}. Although research has adopted the term to describe AI behavior, in our study, humans demonstrate this effect. While this illustrates a major incision in the human knowledge structure, it is crucial for research and practice to develop mechanisms to counteract this effect.

%------impact on task performance-------
\textit{Design guidelines for practice. }Our findings confirm that AI support enhances immediate task performance, even when the explanations are incorrect. Even though this observation aligns with prior HCI studies demonstrating that explanations can increase human-AI team performance, such as those by \citet{bansal2019beyond} and \citet{schemmer2023appropriate}, we extend prior research by revealing the negative impacts of explanations. Our work shows that the \textbf{performance boost provided by AI during the collaboration can be short-term} if the explanations are incorrect, as they negatively impact humans' procedural knowledge and reasoning post-collaboration.
%----based on the reasoning findings
Based on our analyses, we can derive further design implications. Developers of XAI tools should consider how to provide explanations to decision-makers and control for a limited amount of cognitive load infused by the information provided to its users. Our findings also indicate that explanations are necessary to increase trust. However, with the correctness of explanations having impacts on participants procedural knowledge and reasoning, explanations need to be developed and integrated with caution. Additionally, humans with a higher perceived AI usefulness seem to better reason, even for incorrect explanations. This suggests that the design of explanations should be tailored to individuals’ characteristics (e.g., incentivized approaches) in terms of AI usefulness. If participants perceive the AI as useful, they establish higher reasoning abilities. 
%------impact on procedural knowledge development-------
This finding advances the existing literature on XAI, which has primarily emphasized factors like transparency, reliance, and user understanding during the collaboration\cite{lai2019human, ehsan2018rationalization, morrison2024impact}. Our research indicates that while correct explanations foster procedural knowledge, incorrect explanations can lead to the erosion of knowledge over time, even when they temporarily enhance task performance in collaboration with AI.
Yet, we note that these implications have to be taken with caution. Even though we illustrate that incorrect explanations have the potential to mislead humans, we cannot generalize for every decision-making scenario. Depending on the context and domain, the impact of incorrect explanations might vary, and organizations have to implement prevention mechanisms designed for the specific scenarios. 
%------the role of humans deciding autonomously-------
The findings also underscore the importance of designing AI that \textbf{prioritize the correctness of explanations}, especially in contexts where AI support is not always provided and human decision-makers have to be able to decide autonomously, for instance, as required by the EU AI act \cite{EU_AI_Act_2021}. We suggest that AI should integrate mechanisms allowing humans to evaluate the correctness of explanations critically. For instance, incorporating counterfactual explanations, as explored by previous research \cite{wachter2017counterfactual, goyal2019counterfactual, schemmer2023towards}, could help users better understand the AI’s decision-making process and mitigate the adverse effects of incorrect explanations. Similarly, maintaining the human ability to perform tasks autonomously without AI support is favorable at the workplace. The role of XAI can impact the transfer of knowledge to counter risks of demographic change \cite{spitzer2024transferring}. Incorrect explanations impairing knowledge development have the potential to endanger the successful business operations of organizations. Therefore, we advise designers to incorporate not only mechanisms to detect incorrect explanations post-hoc \cite{xu2023combating, zhou2023synthetic, hartwig2023landscape} but also in the design of the AI.

%------automation bias-------
\textit{The impact for organizations.} Prior research has mainly explored effects that occur \textbf{during} the collaboration of humans and AI. For instance, \citet{schemmer2023appropriate} conceptualize the relationship between the appropriateness of reliance and how it relates to the human-AI team performance. \citet{morrison2024impact} advance this view by the dimension of explanations and explore the effects of their correctness on humans' appropriate AI reliance.
By outlining potential downsides of XAI, this work addresses the impact of explanations' correctness on humans \textbf{post-collaboration}. Our study demonstrates that their procedural knowledge and reasoning are impaired when they are provided with incorrect explanations and the AI support is removed. Taking in a human-centric perspective \cite{horvatic2021human}, this repercussion presents harm to not only humans' individual knowledge development \cite{bhatt2000organizing} but also to their ability to provide meaningful assets to organizations \cite{nonaka2007knowledge, davenport1998working}. Maintaining individuals as valuable assets to organizations is crucial because it directly influences organizational innovation, efficiency, and adaptability in a competitive market. \citet{schemmer2021intelligent} emphasize that the design of decision-support systems can encourage automation bias and, consequently, deskill human users. Sustaining humans' knowledge development can foster a more resilient and informed workforce capable of driving sustained success \cite{grant1996prospering}.

By addressing these complex dynamics, this study contributes to the advancement of HCI as a field, offering practical insights for the design of AI that are both effective for human-AI collaboration and beneficial for humans' procedural knowledge development.

\subsection{Limitations and Future Work}
Despite the valuable insights gained from this study, several limitations must be acknowledged, offering avenues for future research. First, the study explores how incorrect explanations affect humans in AI-assisted decision-making. To investigate the impact of such explanations, we designed a study with four treatments of different AI support: no support, only AI advice, AI advice with correct explanations, and AI advice with incorrect explanations. In real-world applications, the interaction with AI that provides only correct or incorrect explanations is rather unlikely. It presents valuable means to take the first steps to investigate the impact of incorrect explanations but does not reflect the real world. Future research could take on this aspect to extend our findings and evaluate how a mix of correct and incorrect explanations---a mix that is more realistic for deployed AI---affects humans' procedural knowledge and reasoning. In particular, different ratios of the correctness of explanations could provide further insights and advance the field.

On top of that, our results might not generalize for every human-AI collaboration scenario. We set up a study and gain empirical insights for a decision-making task in which we deploy an LLM to support humans in classifying architectural styles. Even though this task resembles various real-world use-cases in which generative AI is employed to provide advice and additional reasoning (e.g., patients seeking advice for making an initial medical diagnosis or receiving advice for financial investments), we cannot generalize our findings for every context. In different modalities with other task objectives (e.g., content creation) and conditions (e.g., duration of the study), incorrect explanations might show slightly different effects.

Second, the focus of the study was on measuring task performance to derive insights into humans' procedural knowledge and inform about the impact on human-AI team performance. In addition, we used an LLM-based approach to derive the reasoning abilities of participants. First, such an approach to evaluate the findings poses certain challenges, as LLMs might incorporate biases \cite{chen2024humans, shi2024judging}. Second, in real-world scenarios, performance might not be the only metric relevant. Other measures, like appropriate reliance \cite{schemmer2023appropriate} or fairness \cite{schoeffer2024explanations} in the AI, might also be of high relevance in AI-assisted decision-making as previous studies show \cite{hemmer2023human, bansal2021does}. It is of high importance to explore how these factors change over time and under the effect of incorrect explanations. Exploring the temporal implications can extend the views and offer new insights that support the robust and effective design of AI.

Lastly, the study primarily relied on short-term measures based on task performance, which may not fully capture the long-term impact of AI support on human knowledge and how the correctness of explanations impacts the human-AI team performance. In our work, participants take, on average, 15 minutes to conduct the study. While such an interaction with AI reflects common real-world scenarios where users briefly consult AI systems for immediate decision support (e.g., seeking advice for math equations in homework or requesting a classification for quality checks), it limits our ability to observe longer-term effects. The insights we gained even in this brief interaction suggest that incorrect explanations can rapidly influence human understanding, though future work should examine whether these effects persist or evolve over extended periods of engagement. The short-term nature of our study may actually underestimate the full impact of incorrect explanations, as longer exposure could lead to deeper entrenchment of misunderstandings. Future research could employ longitudinal designs to assess how incorrect explanations influence procedural knowledge development and reasoning over extended periods and across multiple tasks. This approach would offer a deeper understanding of how different types of explanations contribute to sustained knowledge development, aligning with the principles of human-centered AI.

%% file: 07-conclusion.tex
\section{Conclusion}

This work sets out the first steps towards investigating the effect of incorrect explanations on the human and the human-AI team. By doing so, we take a human-centric perspective and analyze the repercussions of incorrect explanations on task performance to derive insights into humans' procedural knowledge and reasoning. In an online study, we assessed the impact of such explanations, specifically after the AI support is withdrawn, and humans must act autonomously. 

With our work, we make several contributions to the HCI field: First, we identify a misinformation effect caused by incorrect explanations, which impairs humans' procedural knowledge and reasoning. Second, we offer insights into how such incorrect explanations limit human-AI team capabilities. Finally, we provide guidelines for the effective and safe design of explanations that can foster AI-assisted decision-making. 
\textit{So we can eventually imagine: the AI provides a correct explanation for differentiating the architectural styles. You pass your exam.}

%% file: 08-appendix.tex
\section{Appendix}
\label{sec: appendix}

\begin{table}[htbp!]
\caption{Architecture styles and their descriptions}
\Description{The table shows the description of the three architecture styles used in the study.}
\label{tab: descriptions}
\centering
\begin{tabular}{m{3cm} m{11cm}} \hline
\textbf{Architecture styles} & \textbf{Descriptions} \\
\hline
Art Nouveau & 
\begin{itemize}
    \item Art Nouveau draws inspiration from the natural world, employing sinuous, sculptural, and organic shapes. The undulating asymmetry of these lines creates a sense of movement and vitality.
    \item The style embraces arches and curving lines. Architects aimed to create something fresh and vital, resulting in buildings with unique and expressive forms.
    \item Art Nouveau architects experimented with modern materials, including iron, glass, ceramics, and later concrete. The movement rejected rigid historical conventions in favor of innovation.
\end{itemize} \\
\hline
Art Deco & 
\begin{itemize}
    \item Many Art Deco facades feature angular, upward-pointing vertical lines. These triangular shapes culminate in a series of steps, creating a sense of upward movement and dynamism.
    \item Art Deco embraces bright, opulent colors. Stark contrasts, such as black and white or gold and silver, create a striking visual impact.
    \item Art Deco buildings seamlessly blend stucco, terra-cotta, decorative glass, chrome, steel, and aluminum. These materials contribute to the sleek yet opulent appearance of structures.
\end{itemize} \\
\hline
Georgian Architecture & 
\begin{itemize}
    \item Georgian buildings are meticulously symmetrical. Their facades exhibit a harmonious arrangement of windows, doors, and other architectural features.
    \item Historic Georgian homes feature doors with fanlights above. Additionally, the doors are set back at least four inches from the brick face.
    \item Georgian buildings favor time-tested materials. Bricks, stone, and wood form the backbone of these structures, emphasizing longevity and solidity.
\end{itemize} \\
\hline\\\\
\end{tabular}
\end{table}

\begin{table}[htbp!]
\caption{Prompts for Correct and Incorrect Explanations and for reasoning scores}
\Description{The table shows the prompts used to generated the output of the LLM.}
\label{tab: prompts}
\centering
\begin{tabular}{m{3cm} m{11.3cm}} \hline
Prompt for correct explanations & \textit{Hello, you have to provide support to a human in classifying the architectural style of buildings in images. Overall, there are three different styles to choose from: Art Deco, Art Nouveau and Georgian Architecture. For an image, please also provide within two to three sentences an explanation for your decision. In the following is the image:} \\
\hline
Prompt for incorrect explanations & \textit{Hello, you have to classify the architectural style of buildings in images. Overall, there are three different styles to choose from: Art Deco, Art Nouveau, and Georgian Architecture. Please provide the classification of an image. Additionally, we want to investigate how false explanations can mislead users. Thus, also explain your decision within 2-3 sentences. However, make sure that the explanation is incorrect, thus it does not correspond to your classification. Can you please provide a misleading explanation by using actual elements of the image? Here is the image:} \\
\hline
Prompt for reasoning scores & \textit{Please Provide an assessment of the correctness of the human explanation, and give the response back with only a correctness score for each human explanation. Here is a human explanation and ground truth explanation regarding three different architectural styles: Art Nouveau, Art Deco, and Georgian Architecture. The human explanation is divided into three parts in the order of Art Nouveau, Art Deco, and Georgian Architecture. Human Explanation: $\{ai_explanations\}$ Ground Truth: $\{ground_truths\}$. All Ground truths can be seen as high-score explanations. Correctness refers to the truthfulness/faithfulness of the explanation. Decide on a scale of 0-5 for scoring explanations: 0: Completely incorrect or irrelevant. 1: Major errors or omissions; very limited correct information. 2: Some correct information, but with significant errors or incomplete coverage. 3: Generally correct but with minor inaccuracies or omissions. 4: Mostly accurate and complete, with minor errors. 5: Completely correct and comprehensive.} \\
\hline\\\\
\end{tabular}
\end{table}

\begin{table}[htbp!]
\caption{Variables and items of the questionnaire}
\Description{The table shows the different variables and items used in the questionnaire of the study.}
\label{tab: items}
\centering
\begin{tabular}{m{2cm} m{12cm}} \hline
\textbf{Variable} & \textbf{Items} \\
\hline
AI knowledge & 
\begin{itemize}
    \item How much experience do you have in working with an AI?
\end{itemize} \\
\hline
AI usability & 
\begin{itemize}
    \item Using the AI system would enable others to classify architectural styles of buildings more quickly.
    \item Using the AI system would improve the performance when classifying architectural styles of buildings.
    \item Using the AI system would increase productivity for classifying the architectural styles of buildings.
    \item Using the AI system would enhance my effectiveness of classifying the architectural styles of buildings.
    \item Using the AI system would make it easier for classifying the architectural styles of buildings.
    \item I would find the AI system useful for classifying the architectural styles of buildings.
    \item Learning to operate the AI would be easy for me.
    \item I would find it easy to get the AI to do what I want it to do.
    \item My interaction with the AI would be clear and understandable.
    \item I would find the AI to be flexible to interact with.
    \item It would be easy for me to become skillful at using the AI.
    \item I would find the AI easy to use.
\end{itemize} \\
\hline
AI trust & 
\begin{itemize}
    \item The AI is deceptive.
    \item I am suspicious of the AI's intent, action, or outputs.
    \item The AI's actions will have a harmful outcome.
    \item I am confident in the AI.
    \item The AI is reliable.
    \item I can trust the AI.
\end{itemize} \\
\hline
Cognitive load & 
\begin{itemize}
    \item How mentally demanding was the task?
    \item How physically demanding was the task?
    \item How hurried or rushed was the pace of the task?
    \item How hard did you have to work to accomplish your level of performance?
    \item How insecure, discouraged, irritated, stressed, and annoyed were you?
\end{itemize} \\
\hline\\\\
\end{tabular}
\end{table}

\begin{table}[htbp!]
\caption{Statistical test results to compare the incorrect explanations treatment with the other treatments for the main task.}
\Description{The table shows the results of the statistical test for the main task to compare differences between the incorrect explanation treatment and the other treatments.}
\label{tab: maintaskperf}
\centering
\begin{threeparttable}
\begin{tabular}{m{6cm} ccc} \hline
\multicolumn{1}{m{6cm}}{Comparison} & t-statistic & p-value & p-value (corrected) \\
\hline \hline
incorrect explanations vs. control group & 5.637  & .999  & 1.000 \\
incorrect explanations vs. AI  & -.546  & .293  & .293 \\
incorrect explanations vs. correct explanations  & -2.359  & .010  & .021 \\
\hline
\end{tabular}
\end{threeparttable}
\end{table}

\begin{table}[htbp!]
\caption{Regression results for the main task performance}
\Description{The table shows the results of the regression model with the main task performance as the dependent variable. In the columns, the coefficient and the standard error are presented. As significance levels, .1, .05, and 0.01 are defined.}
\label{tab: regressionmain}
\begin{threeparttable}
\begin{tabular}{m{7cm} R{1.5cm} R{1.5cm}} \hline
\multicolumn{1}{m{7cm}}{Dependent variable} & \multicolumn{2}{c}{Performance} \\
\cmidrule{2-3}
& Coeff & SE \\
\hline \hline
Intercept                                         & .676***   & .083  \\
condition\_AI [True]                               & .259***   & .039  \\
condition\_correctexplanation [True]               & .311***   & .040  \\
condition\_incorrectexplanation [True]             & .250***   & .038  \\
cognitive\_load                                   & -.023     & .087  \\
AI trust                                         & -.237*    & .134  \\
AI knowledge                                      & -.029     & .061  \\
AI usefulness                                       & .092      & .096  \\ \hline
$R^{2}$                                           & \multicolumn{2}{c}{.367} \\
Adj. $R^{2}$                                           & \multicolumn{2}{c}{.338} \\
Log-Likelihood                                    & \multicolumn{2}{c}{59.892} \\
F-statistic                                       & \multicolumn{2}{c}{12.580***}
\\\hline
\end{tabular}
    \begin{tablenotes}
        \item[1] Note: \textit{*} \textit{p < .1}; \textit{**} \textit{p < .05};  \textit{***} \textit{p < .01}\newline\newline
    \end{tablenotes}
\end{threeparttable}
\end{table}

\begin{table}[htbp!]
\caption{Statistical test results to compare the incorrect explanations treatment with the other treatments for the post-test.}
\Description{The table shows the results of the statistical test for the post-test to compare differences between the incorrect explanation treatment and the other treatments.}
\label{tab: posttestperf}
\centering
\begin{threeparttable}
\begin{tabular}{m{6cm} c c c} \hline
\multicolumn{1}{m{6cm}}{Comparison} & t-statistic & p-value & p-value (corrected) \\
\hline \hline
incorrect explanations vs. control group & 2.241  & 0.986  & 0.986 \\
incorrect explanations vs. AI & -0.371  & 0.356  & 0.356 \\
incorrect explanations vs. correct explanations & -1.742  & 0.043  & 0.085 \\
\hline\\\\
\end{tabular}
\end{threeparttable}
\end{table}

\begin{table}[htbp!]
\caption{Regression results for the post-test performance}
\Description{The table shows the results of the regression model with the post-test performance as the dependent variable. In the columns, the coefficient and the standard error are presented. As significance levels, .1, .05, and 0.01 are defined.}
\label{tab: regressionposttest}
\begin{threeparttable}
\begin{tabular}{m{7cm} R{1.5cm} R{1.5cm}} \hline
\multicolumn{1}{m{7cm}}{Dependent variable} & \multicolumn{2}{c}{Performance} \\
\cmidrule{2-3}
& Coeff & SE \\
\hline \hline
Intercept                                         & .618***   & .112  \\
condition\_AI [True]                               & .125**    & .053  \\
condition\_correctexplanation [True]               & .198***   & .054  \\
condition\_incorrectexplanation [True]             & .117**    & .052  \\
cognitive\_load                                   & -.197*     & .118  \\
AI trust                                         & -.136     & ,180  \\
AI knowledge                                      & .006      & .082  \\
AI usefulness                                       & .033      & .130  \\ \hline
$R^{2}$                                           & \multicolumn{2}{c}{.127} \\
Adj. $R^{2}$                                           & \multicolumn{2}{c}{.087} \\
Log-Likelihood                                    & \multicolumn{2}{c}{11.914} \\
F-statistic                                       & \multicolumn{2}{c}{3.153***}
\\\hline
\end{tabular}
    \begin{tablenotes}
        \item[1] Note: \textit{*} \textit{p < .1}; \textit{**} \textit{p < .05};  \textit{***} \textit{p < .01}\newline\newline
    \end{tablenotes}
\end{threeparttable}
\end{table}

\begin{table}[htbp!]
\caption{Statistical test results to compare the procedural knowledge development between groups}
\Description{The table shows the results of the ANOVA analysis to compare differences between the treatments for the difference in post-test performance and pre-test performance.}
\label{tab: procedurcalknowledgedevelopment}

\centering
\begin{threeparttable}
\begin{tabular}{m{3cm} m{3cm} cccc} \hline

\multicolumn{2}{c}{Comparison} &  &  &  &  \\
\cmidrule{1-2}
Group 1 & Group 2 & Mean Difference & p-adj & Lower & Upper \\
\hline \hline
AI                & Control               & -0.1667 & 0.0185 & -0.3129 & -0.0204   \\
AI                & Correct Explanations  &  0.0000 & 1.0000 & -0.1462 &  0.1462  \\
AI                & Incorrect Explanations& -0.0250 & 0.9707 & -0.1712 &  0.1212  \\
Control           & Correct Explanations  &  0.1667 & 0.0185 &  0.0204 &  0.3129   \\
Control           & Incorrect Explanations&  0.1417 & 0.0613 & -0.0046 &  0.2879  \\
Correct Explanations & Incorrect Explanations & -0.0250 & 0.9707 & -0.1712 &  0.1212 
\\
\hline
\end{tabular}
\end{threeparttable}
\end{table}

\begin{table}[htbp!]
\caption{Exemplary answers to the open-ended questions. We color-highlighted words referring to a correct (\textcolor{ForestGreen}{green}) and incorrect (\textcolor{Maroon}{red}) answer.}
\Description{Randomly selected answers to the open-ended questions.}
\label{tab: reasoninganswers}
\centering
\begin{tabular}{m{1.3cm} m{3.6cm}m{3.6cm}m{3.6cm}m{1.2cm}} \hline
\multicolumn{1}{m{1.3cm}}{Participant} & \multicolumn{3}{c}{Answers} & Reasoning score \\
\cmidrule{2-4}
& Art Nouveau & Art Deco & Georgian \\
\hline \hline
P7 & \textit{Art Nouveau draws \textcolor{ForestGreen}{inspiration} for \textcolor{ForestGreen}{nature} in \textcolor{ForestGreen}{lines} and \textcolor{ForestGreen}{form} while still maintaining \textcolor{ForestGreen}{modernity} and \textcolor{Maroon}{simplicity} in building materials.  The \textcolor{Maroon}{flow} of the lines complement rigid forms and lines, and \textcolor{ForestGreen}{assymetry} is celebrated.} & \textit{Art Deco is all about \textcolor{ForestGreen}{geometry}, \textcolor{ForestGreen}{curve}, and sleek \textcolor{ForestGreen}{modernism} (as interpreted in the early 20th century). The lines are \textcolor{Maroon}{smooth} and \textcolor{Maroon}{fluid}, yet with some \textcolor{ForestGreen}{repetition} that is very pleasing. \textcolor{ForestGreen}{Symmetry} is celebrated and enhanced with that same \textcolor{ForestGreen}{repetition}, and you can get a sense of \textcolor{ForestGreen}{strong} and \textcolor{ForestGreen}{soaring} feelings in some of the buildings and compact \textcolor{ForestGreen}{sleekness} in others.} & \textit{Georgian Architecture is a late \textcolor{ForestGreen}{17th}-early \textcolor{ForestGreen}{18th} century style that celebrates \textcolor{ForestGreen}{symmetry}, \textcolor{ForestGreen}{repetition}, and plain \textcolor{ForestGreen}{lines} sometimes enhanced with masonry features also very \textcolor{ForestGreen}{strong} and \textcolor{ForestGreen}{simple} in their form. \textcolor{ForestGreen}{Brick}, \textcolor{ForestGreen}{stone}, and \textcolor{ForestGreen}{stucco} are very much in evidence. The \textcolor{ForestGreen}{entrance} is a dominant feature, often with a \textcolor{ForestGreen}{fan} \textcolor{ForestGreen}{window} above the \textcolor{ForestGreen}{door} and \textcolor{ForestGreen}{column} or column-like surrounds with \textcolor{ForestGreen}{decorative} \textcolor{ForestGreen}{capitals}.} & 3.77 \\\hline

P34 & \textit{I would define architectural style Art Nouveau as the form of architecture that incorporate \textcolor{ForestGreen}{natural} and \textcolor{Maroon}{historic} features in environment in its \textcolor{ForestGreen}{design}.} & \textit{Architecturral style Art Deco is the form of architecture that incorporates a lot of \textcolor{ForestGreen}{angular} and \textcolor{ForestGreen}{triangular} designs and are also very \textcolor{Maroon}{colourful}.} & \textit{Georgian Architecture is a form of architecture that incorporates a lot of windows and doors in its design. These \textcolor{Maroon}{door} and \textcolor{Maroon}{windows} are fixed at a \textcolor{Maroon}{distance} from the \textcolor{Maroon}{wall}.} & 2.0\\\hline

P74 & \textit{\textcolor{ForestGreen}{flowing}, \textcolor{ForestGreen}{organic} shapes inspiMaroon by \textcolor{ForestGreen}{nature}, \textcolor{ForestGreen}{curving} lines, \textcolor{ForestGreen}{assymetry}, and the use of \textcolor{ForestGreen}{materials} like \textcolor{ForestGreen}{iron}, \textcolor{ForestGreen}{glass}, and \textcolor{ForestGreen}{ceramics}. The \textcolor{ForestGreen}{design} often includes \textcolor{ForestGreen}{decorative}, whimsical elements.} & \textit{\textcolor{ForestGreen}{angular}, \textcolor{ForestGreen}{upward-pointing} lines, \textcolor{ForestGreen}{geometric} shapes, and a combination of opulent materials like \textcolor{ForestGreen}{stucco}, \textcolor{ForestGreen}{terra-cotta}, and \textcolor{ForestGreen}{metals}. The style often features \textcolor{ForestGreen}{bold} \textcolor{ForestGreen}{colors} and \textcolor{ForestGreen}{sharp} \textcolor{ForestGreen}{contrasts}.} & \textit{\textcolor{ForestGreen}{symmetrical} facades with a \textcolor{ForestGreen}{balanced} arrangement of \textcolor{ForestGreen}{windows} and \textcolor{ForestGreen}{doors}. Features include \textcolor{ForestGreen}{fanlights} above \textcolor{ForestGreen}{doors} and materials like \textcolor{ForestGreen}{brick}, \textcolor{ForestGreen}{stone}, and \textcolor{ForestGreen}{wood}. \textcolor{ForestGreen}{Georgian} buildings are known for their \textcolor{ForestGreen}{classical} proportions and solid, enduring \textcolor{ForestGreen}{construction}.} & 4.33
\\\hline

P135 & \textit{Art Nouveau seems the most modern of the 3 and can take \textcolor{Maroon}{various} somewhat \textcolor{Maroon}{different} looks.} & \textit{Art Deco seems very \textcolor{Maroon}{old} style. With \textcolor{Maroon}{concrete} buildings that are somewhat \textcolor{ForestGreen}{tall}} & \textit{Georgian Architecture has a specific look that seems to have \textcolor{ForestGreen}{bricks} very often and buildings that do not look \textcolor{Maroon}{modern} at all.} & 1.66 \\
\hline

\end{tabular}
\end{table}

\begin{table}[htbp!]
\caption{Regression results for participants' reasoning.}
\Description{The table shows the results of the regression model with the reasoning results as the dependent variable. In the columns, the coefficient and the standard error are presented. As significance levels, .1, .05, and 0.01 are defined.}
\label{tab: regressionreasoning}
\begin{threeparttable}
\begin{tabular}{m{7cm} R{1.5cm} R{1.5cm}} \hline
\multicolumn{1}{m{7cm}}{Dependent variable} & \multicolumn{2}{c}{Reasoning scores} \\
\cmidrule{2-3}
& Coeff & SE \\
\hline \hline
Intercept                                         & 2.611***   & .349  \\
condition\_AI [True]                               & .171       & .165  \\
condition\_correctexplanation [True]               & .058       & .169  \\
condition\_incorrectexplanation [True]             & -.144      & .161  \\
cognitive\_load                                   & -.109      & .367  \\
AI trust                                         & -.712      & .561  \\
AI knowledge                                      & -.192      & .257  \\
AI usefulness                                       & .744*       & .404  \\ \hline
$R^{2}$                                           & \multicolumn{2}{c}{.066} \\
Adj. $R^{2}$                                      & \multicolumn{2}{c}{.023} \\
Log-Likelihood                                    & \multicolumn{2}{c}{-169.82} \\
F-statistic                                       & \multicolumn{2}{c}{1.540} \\
\hline
\end{tabular}
    \begin{tablenotes}
        \item[1] Note: \textit{*} \textit{p < .1}; \textit{**} \textit{p < .05};  \textit{***} \textit{p < .01}\newline\newline
    \end{tablenotes}
\end{threeparttable}
\end{table}